\documentclass{article}

\usepackage{mathptmx}
\usepackage{helvet}
\usepackage{courier}
\usepackage{type1cm}         

\usepackage{makeidx}         % allows index generation
\usepackage{graphicx}        % standard LaTeX graphics tool
\usepackage{multicol}        % used for the two-column index
\usepackage[bottom]{footmisc}% places footnotes at page bottom

\begin{document}
 \bibliographystyle{spmpsci}

\author{Luisa Bonolis}
\title{Bruno Rossi and Cosmic Rays: From Earth laboratories to Physics in Space}

\maketitle

\section*{Abstract}
Rossi's career paralleled the evolution of cosmic-ray physics. Starting from the early 1930s his pioneering work on the nature and behavior of cosmic rays led to fundamental   contributions in the field of experimental cosmic-ray physics and laid the foundation for high-energy particle physics. After the war, under his leadership the Cosmic Ray group at MIT investigated the properties of the primary cosmic rays elucidating the processes involved in their propagation through the atmosphere, and measuring the unstable particles generated in the interactions with matter. When  accelerators came to dominate particle physics, Rossi's attention focused on  the new opportunities for exploratory investigations made possible by  the availability of space vehicles. He initiated a research program which led to the first {\it in situ} measurements of the density, speed and direction of the solar wind at the boundary of Earth's magnetosphere and  inspired  the search for extra-solar X-ray sources resulting in the detection of what revealed to be the most powerful X-ray source in Earth's skies. The discovery of Scorpius X-1 marked the beginning of X-ray astronomy, which soon became a principal tool of astrophysics research.

\section{Introduction}
%Bruno Rossi  opened the Preface of his 1990 autobiography, {\it Moments in the Life of a Scientist} \cite{Rossi:1990aa}, with the following words: 
%\begin{quotation}
%\small

%At this time, when recent developments have brought science to a prominent position both in the cultural life and in the everyday life of our society, it is of interest to recall the activities of scientists in the years when the foundations of these truly revolutionary developments were laid. I was one of these scientists.''
%\end{quotation}
%\normalsize

 Between the beginning of 1930 and the fall of 1938, a period full of ``ironies of wonder and of fear,''\footnote{Philip Morrison, ``Foreword'' in \cite{Rossi:1990aa}, p. xi.} Bruno Rossi became an internationally known scientist as one of the leading actors in the study of the nature and behavior of cosmic radiation. His life and scientific career were tightly entangled with earthshaking historical events which turned upside down the world panorama.\footnote{The main biographical information on Bruno Rossi is contained in his autobiography \cite{Rossi:1990aa}, in a short autobiographical note \cite{Rossi:1974fk}, and in some historical essays containing personal recollections \cite{Rossi:1981aa}  \cite{Rossi:1982aa}  \cite{Rossi:1983uq}  \cite{Rossi:1984fk} \cite{Rossi:1985fk}.  Further news on his life are present in George Clark's excellent biographical notes published on the \textit{Proceedings of the American Philosophical Society}  \cite{Clark:2000fk} as well as in interviews such as those preserved at the American Institute of Physics and at Los Alamos Archives\index{Los Alamos!Archives}  or in private collections.}
%AGGIUNGERE CONTRIBUTO DI CLARK SU BIOGRAFICO

 In parallel with his great contributions to modern science, Rossi, who was the author of books on ionization chambers and counters, optics, cosmic rays, high-energy particles and space physics \cite{Rossi:1949fk}  \cite{Rossi:1957eu}  \cite{Rossi:1966aa}  \cite{Rossi:1952aa}  \cite{Rossi:1970uq},  formed new generations of talented researchers in physics and astrophysics, and was a source of inspiration and a guide for anyone who came in contact with him. This is also testified by the great deal of personal recollections, that bear witness to the impact of his giant human and scientific personality.\footnote{See in particular the following writings published mainly on commemorative occasions: \cite{Puppi:1995fk}  \cite{Pacini:1995aa}  \cite{Scarsi:1995aa}  \cite{Cocconi:1995aa}  \cite{Morrison:1997aa}  \cite{Bonetti:1997zm}  
\cite{Clark:2000fk}  \cite{Bonolis:2002aa}   \cite{Scarsi:2005aa}  \cite{Palma:2005aa}  \cite{Clark:2006aa}  \cite{Bonetti:2006ab}  \cite{Scarsi:2006aa}  \cite{Canizares:2006aa}.} %Obituaries Aggiungi Morrison altro saggio

Rossi's scientific life is extending across a period of about thirty years from the end of the 1920s  to the early 1960s. The ``Italian Years'' include Bruno Rossi's scientific activity from the beginning of his academic and scientific life in Florence, in 1928, until 1938, when he was dismissed from his chair in Padova and was forced to emigrate  by fascist racial laws  \cite{Bonolis:2011fk}.
%\footnote{About 60 papers were written by Rossi during the period 1928--1938.} 
 At the beginning of 1930s he was a pioneer of cosmic ray research in Italy, while Enrico Fermi and Franco Rasetti were building up a research group in Rome, the group who would share with them the well known achievements in nuclear physics. 

Unlike the ``Roman School'' the Florence group was rather an informal community of young physicists, sharing an enthusiasm in searching for new lines of research under the protective wings of Antonio Garbasso\index{Garbasso, Antonio}, the Director of the Physics Institute.  It is to be remarked how Rossi and the other young people of the Florentine group were at the very beginning of the 1930s at the forefront of the experimental research in Italy, both regarding the choice of the subject as well as the techniques they employed.  After introducing the brand new Geiger-M\"uller counter in Italy \cite{Rossi:1930gf}, Rossi immediately devised the coincidence circuit, a device made of Geiger-M\" uller counters and electronic valves (the AND logic circuit) capable of registering the simultaneous occurrence of electrical pulses from any number of counters \cite{Rossi:1930mz}. Rossi's innovative techniques were soon used by Gilberto Bernardini\index{Bernardini, Gilberto} and Daria Bocciarelli\index{Bocciarelli, Daria} in experiments on what we might name the nuclear physics of the time, thus someway anticipating the Rome group, who was more or less still working at an ``atomic level'',  with some exceptions, notably Rasetti's\index{Rasetti, Franco} relevant researches on the Raman effect, and some works on the hyperfine structure of spectra, caused by nuclear effects. The Geiger-M\"uller counters were also the basis for the magnetic spectrograph built by Giuseppe Occhialini\index{Occhialini, Giuseppe Stanislao} to catch the beta rays emitted by weakly radioactive substances \cite{Occhialini:1931uq}. Apparently the Florentine group was working within a much freer environment, where no such ``dominating'' figures as Fermi\index{Fermi, Enrico}  and Rasetti\index{Rasetti, Franco} could influence, in a way or another, their activity. Actually, the skill and experience of the Arcetri Group in the art of  building Geiger-M\"uller tubes and coincidence circuits was instrumental in 1934, when neutron work began in via Panisperna \cite{Leone:2005aa}. It became one of the roots of the well known achievements and discoveries of the Roman Group \cite{Amaldi:1984uq}, for which Fermi\index{Fermi, Enrico}  was awarded the Nobel Prize in 1938.

The racial laws as well as personal opposition to fascist regime forced relevant figures such as Fermi, Rossi, Rasetti, Giulio Racah, Emilio Segr\`e, Giuseppe Occhialini, to leave Italy \cite{Orlando:1998jt}. This brain-drain was at the root of what Edoardo Amaldi\index{Amaldi, Edoardo} dubbed ``the disaster of Physics in Italy'' \cite[p. 186]{Amaldi:1979de}.   However, during the 1920s and 1930s  these founders of  modern physics had laid down a basis which after World War II would definitely show its vitality. Rossi's young colleague Gilberto Bernardini\index{Bernardini, Gilberto} cooperated with Edoardo Amaldi\index{Amaldi, Edoardo},  Fermi's former student, in the reconstruction of Italian physics since the difficult war years and both had a key role in the the building of a brand new reality during the 1950s.

Despite the relevance of Rossi's impact on 20th century physics, and his importance also as a teacher and a promoter of scientific research during his whole life, which went through most dramatic events of the past century, Rossi's scientific work, and his leading role has not attracted the attention of historians of physics as it deserves. The Italian scientific community, excepted a few physicists, has the tendency to generically identify Rossi with cosmic-ray research,  someway ``forgetting'' his later pioneering contribution to the field of physics in space, as well as his role as a promoter of the birth of X-ray astronomy, probably because these achievements were on one side strongly connected with his activity in the United States, but also because his contribution to these beginnings were obscured by the later giant developments in the field. 

Rossi's relevant role in the history of modern physics in Italy aroused a due interest in Italian historians, who, starting from the 1980s, began  to study the initial phase of Rossi's investigations, duly focusing on his pioneering role in the cosmic raysÕ physics in 1930s, and emphasizing the birth of the research group of Arcetri \cite{Mando:1986ly}  \cite{Bonetti:2006aa}  and \cite{BonettiMazzoni:2007} in the same years Fermi\index{Fermi, Enrico}  was building up in Rome the group later known as the ``boys from via Panisperna''. Outstanding pioneering work has thus been done on Rossi's  ``Italian period'' (late 1920s to 1938), particularly on the most manifest aspects of his early activity in the 1930s. It has been outlined the paramount relevance of Rossi's invention of the coincidence circuit, and his prediction and independent discovery of the geomagnetic East-West effect \cite{DAgostino:1984fk} \cite{Russo:1988aa} \cite{De-Maria:1989aa} \cite{De-Maria:1992aa} \cite{Bustamante:1994aa} \cite{De-Maria:1994aa} \cite{Russo:2000aa} \cite{Russo:2000ab}  \cite{Peruzzi:2006aa}. In particular, it has been duly emphasized how Rossi introduced the Geiger counter in Italy, the instrument which would play a crucial role in Fermi's\index{Fermi, Enrico}  discovery of the artificial radioactivity induced by neutrons \cite{Leone:2005aa}. 

A partial account of the beginning of Rossi's scientific activity can be also found in his  volume \textit{Cosmic Rays}  \cite{Rossi:1966aa} as well as in some essays about the early period of cosmic ray researches, which testify his interest in history of physics as well as the importance of personal recollections as an unvaluable source of historical information  \cite{Rossi:1966aa}  \cite{Rossi:1981aa} \cite{Rossi:1982aa}.

Besides the aforementioned studies,  Peter Galison's\index{Galison, Peter} relevant analysis of   significant aspects in Rossi's work within a wider perspective is to be duly stressed. In particular in connection with his discussion on the role of cosmic rays in the theoretical and experimental panorama of the developing Quantum Electrodynamics (QED), and of the early history of elementary particle physics \cite{Galison:1987kx} \cite{Galison:1997rm}. 

Even if the ``Italian period'' has already been considered from the point of view of historical research, a more detailed  analysis of Rossi's early studies on the nature and behavior of the cosmic radiation, shows how Rossi's activity is to be considered in its entirety, in order to really clarify his role as a scientist at the very forefront of research. Each single contribution taken into consideration by historians is in fact to be  regarded as part of a much broader and well defined research project. At a time when the only known particles were still the proton and the electron, in studying the interaction between cosmic rays and matter Rossi became a ``particle hunter {\it ante litteram}''.   Furthermore, Rossi's ``Italian years'' are  to be interpreted as an important segment of a longer scientific path during which Rossi showed a strong coherence and adherence to a much wider and visionary plan as a scientist. 
%DISCORSO SULLE CARTE 

The whole arc of Bruno Rossi's life as a scientist, in being so interlaced with 20th century history, went through many changes determined by the dramatic growth and striking developments  which physics underwent during the past century. In this regard, it is  in fact to be remarked that all historical research articles on his scientific activity do not tackle the long period starting from Rossi's emigration from Italy in 1938, to his involvement in war research, and the beginning of his activity at Massachusetts Institute of Technology when he founded the Cosmic-Ray Group which had a leading role during the end of the 1940s and beginning of 1950s, when cosmic rays were still the main source for high-energy particles.\footnote{An excellent view of Rossi's activity  related to his ``American period'' is presented in the already cited biographical notes prepared by his former collaborator George W. Clark \cite{Clark:2000fk} \cite{Clark:2006aa}.}  
%CONTROLLARE CONTRIBUTI DI CLARK E AGGIUNGERE BIO ULTIMA SUL BIOGRAFICO
The advent of high-energy accelerators during the 1950s provided artificial beams of particles which began to challenge cosmic rays as a source of energetic particles, even if theory was far from being able to give an overall explanation for the properties of the increasing number of new particles which were still being discovered by cosmic-ray physicists. At the time many of them chose to work with machines. This was not Rossi's case, who was not attracted by this way of doing physics, even if he was  a leading actor in particle physics research of those years, and for a long time a consultant for the Brookhaven accelerators. When the study of extensive air showers generated by high-energy particles in cosmic rays ---energies already very far from accelerators' possibilities--- gradually changed the perspective of this kind of research, Rossi's astrophysics interests increased and he turned more and more his attention  to outer space and later to  the new possibilities offered by the beginning of space era. 

The period going from the late 1950s to the beginning of 1960s, was the time for the last change of direction in RossiÕs ``scientific odyssey'' \cite[p. 22]{Clark:2000fk}. A change largely determined by a new turn in the technological opportunity offered by space crafts. He certainly was a protagonist of the nascent space age, both as a high level consultant, as an organizer, and as the conceiver of new lines of investigation, which proved of the utmost importance for the future of various areas of  scientific research. 

To fully understand the grounds of this turning point in Rossi's scientific life, it is necessary to follow some main stages of a process which developed through the whole 1950s. This particular period is also interesting from the point of view of inquiring about the relationship between science and society, being strongly related to the political and historical events typical of the 1950s, and to the advent of \textit{big science} and the beginning of the space age. This scientific transition was also a sort of legacy of the International Geophysical Year (IGY, 1957--1958), a scientific enterprise which was shaped by the Cold War and which led to important discoveries in the earth sciences from the ocean depths to the upper atmosphere, generating a lively interest in the potential of satellites for scientific research.

However, the reconstruction of the gradual shifting of Rossi's attention during the 1950s from Earth as a great laboratory for cosmic ray research, towards the outer space, and, in a sense, towards the whole Universe, as well as his intuition on the importance of opening new technological windows on this new exciting laboratory,  became part of a wider  underlying theme which was instrumental in shaping the evolution of Rossi's scientific identity  from  a ``cosmic-ray physicist'' to a ``cosmic-ray astronomer''.  The following discussion is aiming at throwing some light on the more general reasons of the choice made by some members of the cosmic ray community to retain their identity of \textit{cosmiciens} even facing the ``temptation'' offered by the transition from cosmic ray studies of elementary particles, to the study of particles produced by accelerators. In following this path they changed their ``skin''  several times, and got an ``astrophysical'' soul, which cosmic rays physicists still retain in facing challenges on a larger and larger scale, as a  further striking demonstration of Rossi's scientific far-sightedness.

%\section{Searching for the ``secrets of nature''}

\section{Cosmic rays as a new branch of physics}

Bruno Rossi was born in Venice in 1905. After studying at the University of Padua, he received his doctorate in physics at the University of Bologna in November 1927, and started his academic career at the University of Florence as assistant professor of Antonio Garbasso, the director of the Physics Institute. 

At the end of 1920s particle physics was on the verge of emerging ``out of the turbulent confluence of  three initially distinct bodies of research: nuclear physics, cosmic-ray studies, and quantum field theory,'' as Brown and Hoddeson remarked in their introduction to the proceedings of a symposium dedicated to the birth of particle physics  \cite{Brown-Hoddeson:1983uq}.

 The pioneering experiments carried out by Bothe\index{Bothe, Walther} and Kolh\"orster\index{Kolh\"orster, Werner} during the period 1928--1929 \cite{Bothe:1928vn} \cite{Bothe:1929rt} ---following the advent of the Geiger-M\"uller counter  \cite{Geiger:1928kx}--- marked the end of the first period of cosmic-ray research.  The  ``radiation from above'' had been studied first as a phenomenon linked to atmospheric electricity, and then as a cosmic and geo-physical phenomenon. During the 1920s the main preoccupation of scientists working on cosmic rays was to establish their extraterrestrial origin, an hypotheis which had been put forward basing on experiments carried on in Europe before World War I.\footnote{During the spring and summer of 1912 Victor Hess made seven balloon ascents to heights up to 5300 m concluding that ``a radiation of very high penetration power enters our atmosphere from above'' \cite{Hess:1912fj}. In 1913--1914 Werner Kolh\"orster ascended to an altitude of 9200 m and confirmed Hess' results finding that the air ionization had increased up to ten times its value at sea level \cite{Kolhorster:1913uq}. These experiments led to the hypothesis that part of the ionization found at sea level must be due to radiation of extraterrestrial origin for which Hess coined the name {\it H\"ohenstrahlung} (radiation from above). During 1910 and 1911 measurements of the intensity of the radiation were also independently made  by Domenico Pacini at the Regio Ufficio Centrale di Meteorologia e Geodinamica in Rome, Italy. Pacini enclosed his electroscope in a copper box and immersed it in the Tyrrhenian Sea near Livorno, and in the Bracciano Lake near Rome, measuring a significant decrease of the radiation compared to the surface of the water. He concluded that ``a sizable cause of ionization exists in the atmosphere, originating from penetrating radiation, independent of the direct action of radioactive substances in the soil''  \cite[p. 100]{Pacini:1912ys}. A commented translation of this article made by A. De Angelis is available at arxiv.org/abs/1002.1810; for a discussion on Pacini's forgotten contributions see  \cite{De-Angelis:2010fk} \cite{Carlson:2010uq}.} The variation of cosmic-ray intensity with altitude, their absorption and dissipation, and even their cosmic origin were of interest. Still there was the problem of the nature of cosmic rays, which did not attract general attention, probably because of the widespread belief that the astonishingly penetrating cosmic rays could not be anything else but $\gamma$-rays of very high energy.

Bothe and Kolh\"orster's landmark paper published in the fall of 1929  was based on the method of coincidences between two superimposed Geiger-M\"uller counters and had led the two physicists to conclude that the radiation traversing the two counters within a fixed time coincidences could not have the nature of a secondary radiation resulting from the Compton effect generated by the ultra-$\gamma$-radiation \cite{Bothe:1929rt}. They reported on their 
measurements of the absorption of those hypothetical ``secondary electrons'' by recording the coincidences between two superimposed Geiger-M\"uller counters interleaved with metal plates of increasing thickness. From this arrangement they argued that coincidences could be produced only by individual ionizing particles crossing both counters. 

Bothe and Kolh\"orster's landmark paper which contributed to focus the physicists' interest on the local radiation, i.e. the radiation found at the place where measurements were made, aroused the curiosity of the young Bruno Rossi: ``[it] came like a flash of light revealing the existence of an unsuspected world, full of mysteries, which no one had yet begun to explore. It soon became my overwhelming ambition to participate in the exploration." \cite[p. 43]{Rossi:1966aa}. Rossi quickly realized that this kind of  physics could be performed in a terrestrial laboratory and that the Geiger-M\"uller counter could become the key to open ``a window upon a new, unknown territory, with unlimited opportunities for explorations" \cite[p. 35]{Rossi:1981aa}. The opening of ``new windows on the universe'' would be in fact a \textit{leitmotif} in  Rossi's scientific life. A new era was beginning during which the Geiger-M\"uller counter was for a 
long time the keystone of cosmic-ray physics. Rossi's own life as a scientist became intertwined with all its remarkable developments and applications. Excited and full of enthusiasm for the possibilities raised by the Bothe and Kolh\"orster's experiment, Rossi immediately set to work with the help of his students, 
particularly Giuseppe Occhialini and Daria Bocciarelli, and his colleague Gilberto 
Bernardini. Rossi knew that he had very limited means at his disposal, for this reason the 
novelty of the research topic, and the low cost of the research tools were the key ingredients  of his excitement. He was 24 years old when ``one of the most exhilarating periods'' of his life began \cite[p. 10]{Rossi:1990aa}.
In just a few months he built his own Geiger-M\"uller counters, devised a new 
method for recording coincidences, and began some experiments. 

According to Occhialini, the Geiger-M\"uller counter ``was like the Colt pistol in the American frontier: the `great leveler,' 
namely an inexpensive detector, within the reach of every small laboratory, requiring only good 
scientific intuition and experimental skill to get useful results in the new fields of nuclear 
physics and cosmic rays" \cite[p. 253]{De-Maria:1985uq}. However, whereas the invention of the Geiger-M\"uller counter  represented the opening of ``a new technological window'', the setting up of the coincidence circuit,  a real {\it telescope} for cosmic rays, constituted a giant leap forward, turning into what it is to be considered the very {\it first act} of Rossi's research program \cite{Rossi:1930mz}.\footnote{On these issues see L. Bonolis, Walther Bothe and Bruno Rossi: the birth and development of coincidence methods in cosmic-ray physics,  July 2011 [http://arxiv.org/abs/1106.1365].} If the Geiger-M\"uller was an instrument of precision ---particularly as regards directional effects--- being a tool more discriminating than the ionization chamber previously used to study the penetrating power of cosmic rays,  Rossi's version of the coincidence technique obtained connecting Geiger-M\"uller counters and valves in a $n$-fold ``coincidence circuit'' opened new possibilities of investigation and was fundamental in changing the general thinking in cosmic rays, also becoming a powerful tool in nuclear physics research. In designing his innovative device, which registered {\it only particles traversing all   the counters in whatever geometrical configuration}, Rossi was laying the basis for the future experimental practice in the field.

In retrospect, the rapidity with which Rossi launched into experimental physics and performed outstanding research in a new field, is remarkable. As a beginner, Rossi had to compete with wily old foxes who well knew their 
job as experimenters with a solid background in physics. Also, they had worked for a long time in research institutions characterized by an 
outstanding theoretical and experimental tradition such as those in Germany. Within a few weeks the 
first counter was in operation, and Rossi could now tackle the coincidence technique, 
which was at the core of the Bothe-Kolh\"orster experiment, and with incredible insight and skill 
succeeded in fully developing the capabilities of the method.

 Bothe and Kolh\"orster's\index{Kolh\"orster, Werner} pioneering article ---and Rossi's very early researches--- were instrumental in transforming the field in a branch of modern physics:  for the first time, the physical nature of cosmic rays became accessible to experimentation and cosmic rays became the very object of research.

In short time the young Italian physicist acquired a very high expertise, and  was able to discuss fundamental issues with the older physicists of his time, like Robert Millikan\index{Millikan, Robert A.} and Walter Bothe\index{Bothe, Walther}. The former argued  the primary radiation to be $\gamma$-rays  generated in the process of formation of light  elements in the depths of interstellar space.    In becoming soon aware of the complexity of the observed phenomena, Rossi was neither satisfied with  Millikan's views,  nor  with  Bothe and Kolh\"orster's\index{Kolh\"orster, Werner} explanation of the corpuscular nature of cosmic rays, affirming that the primary  radiation must probably be ``high-energy electrons'' traversing the whole atmosphere only because it did not possess the feature of a secondary radiation generated by what Millikan claimed to be primary $\gamma$-rays. On the occasion of the international conference organized in Rome in 1931 by Fermi and Orso Mario Corbino, Rossi was invited  by Fermi to give an introductory speech on the problem of cosmic rays. In contrast with Millikan's ideas,\footnote{Millikan never forgave Rossi for having explicitly expressed the reasons why he thought that Millikan's\index{Millikan, Robert A.} assumption, according to which cosmic rays must be born from the synthesis of light elements in the Universe, could not be correct.}  his words convey the sense of how, after Bothe and Kolh\"orster's experiment, the cosmic ray problem was still to be considered wide open \cite[p. 51]{Rossi:1931rr}:
\begin{quotation}
\small
The main problem on the nature and origin of the penetrating radiation remains still unsolved. The most recent experiments have produced evidence of such strange events that we are led to ask ourselves whether the cosmic radiation is not something fundamentally different from all other known radiations; or, at least, whether in the transition from the energies which come into play in radioactive phenomena to the energies which come into play in cosmic-ray phenomena the behavior of particles and photons does not change much more drastically than until now it was possible to believe.
\end{quotation}
\normalsize

\section{``Image and logic''}
In following Bothe's intuition on the importance of coincidence methods \cite{Rossi:1931ul}, Rossi's style of work is to be highlighted as exemplary in establishing the ``logic'' research tradition \cite{Galison:1987kx} which paved the way for future investigations. By wisely arranging metal screens and circuits of counters according to different geometrical configurations, Rossi was enacting a definite project which since the very first moment aimed at demonstrating the corpuscular nature of cosmic rays, opposing the theory that considered them as high frequency $\gamma$-rays. The natural development of his program was to go deeper and deeper in investigating the  behavior of what he was quite convinced to be corpuscles, particularly studying their interaction with  matter. Following a definite sequence, Rossi tackled the problem from various perspectives, testing the hypothesis through a series of experimental trials. 

Besides analyzing the corpuscles' reaction in presence of an electromagnet \cite{Rossi:1930ly} \cite{Rossi:1931lq}, he used the Earth magnetic field as a spectrometer \cite{Rossi:1931pd}. If primary cosmic rays were charged particles, they would also be affected by the geomagnetic field before entering into the terrestrial atmosphere. A {\it latitude effect} was actually expected indicating a lower intensity of cosmic rays near the equator, where the horizontal component of the geomagnetic field is stronger, but evidence from experiments did not convince the scientific community.\footnote{A slight variation of the intensity had been observed by Jacob Clay since 1927, resulting from experiments made carrying ionization detectors onboard ships that traversed an extensive latitude range stretching from Netherland to Java \cite{Clay:1927fk} \cite{Clay:1928fk}. But the observed effect could possibly be attributed to the different meteorological conditions present in such different locations. Negative results had been found by Bothe and Kolh\"orster \cite{Bothe:1930uq}, and by Millikan  and G. H. Cameron \cite{Millikan:1928uq} during recent experiences carried out respectively between Hamburg and Spitzbergen and between Bolivia and Canada.}

On July 3, 1930 Rossi submitted a paper in which he conjectured the existence of an {\it East-West effect}\index{East-West effect}, a second geomagnetic effect which would be revealed by an  asymmetry of the cosmic-ray intensity with respect to the plane of the geomagnetic meridian,  with more particles coming from East or West, depending  on whether the particle charge was negative or positive \cite{Rossi:1930ve}. 

In parallel with this line of research, Rossi explored also the interaction between particles and matter. The works published during the first three years of activity in the field of cosmic rays, witness the peculiarity of his approach. On the grounds of his views about the corpuscular nature of cosmic rays, Rossi used his powerful device which allowed the electronic counting  of single particles and entered the ``open window on the surrounding world''. Leaving aside the cosmological and planetary scenario of the first decades of the 20th century, Rossi wondered: ``{\it What is the nature of this radiation}?'' And in so doing, he was definitely entering the field of microscopic physics. 
In order to reveal the interaction between radiation and matter Rossi made a ``logical analysis'' of the  phenomenology  observed in relation to  different configurations of counters and metal screens. Both the metal screens the particles have to go through ---according to the chosen configuration--- as well as the counters and valves,  appear to be part of a logic ``super  net''. In  setting his ``traps'' for particles whose identity was still unknown, Rossi is trying to enter into a direct dialogue with the mysterious ``radiation''.  His strategy, which he openly revealed since the very first articles, was to select a ``beam''  of particles and verify what happened when it went through an alternate sequence of counters and screens. His aim was to verify the existence of concomitant phenomena which would lead to conclusions on the activity of cosmic radiation at the sea level. As stated above, though unaware of it, Bruno Rossi was acting as a ``particle hunter''. Indeed, he used some standard procedures for future particle physicists, i.e. having a beam colliding with  a target connected with detectors and observing what happened, wandering what the nature of the new-born corpuscles might be, making hypotheses on their interactions with matter. 

The recording techniques which had been used to reveal  the presence of electrons and $\alpha$ particles since Rutherford times, and the first coincidence techniques used by Walther Bothe\index{Bothe, Walther} to demonstrate the conservation of energy in the collision between electron and photon in the Compton process \cite{Bothe:1924lq} \cite{Bothe:1925fk} \cite{Bothe:1925vn},  had both been applied in a  somehow defined realm, at least by the knowledge of the identity of the microscopic objects whose activity, however, had to be yet understood. On the contrary, the case of cosmic rays regarded the nature itself of the microscopic objects and it thus constituted for some time the main objective of the research which for many years continued to be carried in terrestrial laboratories. At the moment  problems regarding the primary radiation, had to be left aside; its study required the possibility of reaching the highest  layers of earth's atmosphere which only later would become available.

Rossi's initial contribution between 1930 and 1933 has to be compared with the first researches on the cosmic rays prior to his involvement, in order to understand the relationship between cosmic ray research as a new branch of physics and the foundation of the new field of particle physics. His early researches must be in fact examined on the background of a wider perspective regarding the study of the interaction between radiation, particles and matter, a problem strictly intermingled with the debate over the beta decay, the behavior of electrons as beta particles, as well as the difficulty in considering the nucleus as a quanto-mechanical system made of protons and electrons.  Cosmic rays, freely delivered by nature, appeared in fact to be highly relevant  to fundamental problems of physics. 

Rossi's investigations pointed to the existence of two components in cosmic rays
at sea level: a ``hard" component, able to pass through 1\,m of lead
after being filtered through a 10\,cm thick lead screen \cite{Rossi:1932cr}, and a ``soft" component, generated in the atmosphere by primary cosmic rays and able to
generate groups of particles in a metal screen before being stopped \cite{Rossi:1932fk} \cite{Rossi:1933uq}. Rossi's experiments showed that hard and soft rays 
were fundamentally different in character and did not differ merely in their energy. Such unpredicted behavior of the radiation found its confirming synthesis in a curve which was to be universally known as the {\it Rossi curve\index{Rossi curve}} \cite{Rossi:1932qy} \cite{Rossi:1933uq}.

Actually, Rossi's early experiments represented a key contribution along the path leading to the understanding of the role involving electrons and photons  in the formation of the electromagnetic cascade, even if at first they were not fully appreciated or even understood, mainly because of the lack of theoretical tools, as well as of knowledge on particles and fields. The novelty of those results can also be measured by the reaction of the journal \textit{Die Naturwissenschaften} rejecting the article; indeed, it was Werner Heisenberg\index{Heisenberg, Werner K.} to vouch for its credibility, providing its publication in the {\it Physikalische Zeitschrift} \cite{Rossi:1932fk}. 

On the other side, Millikan\index{Millikan, Robert A.} had explicitly attacked the ``so-called Geiger counter coincidence measurements'': ``I have been pointing out for two years in Pasadena seminars, in the Rome Congress on nuclear physics in October, 1931, in New Orleans last Christmas at the A.A.A.S. meeting, and in the report for the Paris Electrical Congress, that these counter experiments never in my judgment actually measure the absorption coefficients of anything. I shall presently show that no appreciable number of these observed ionizing particles ever go through more than 30 cm or at most 60 cm of lead   and yet both Regener\index{Regener, Erich} and Cameron\index{Cameron, G. Harvey} and I have proved that the cosmic rays penetrate through the equivalent of more than 20 feet of lead. These figures cannot both be correct without carrying with them the conclusion that \textit{the primary rays at sea level and below are not charged particles} [emphasis added]'' \cite[p. 663]{Millikan:1933uq}. Millikan\index{Millikan, Robert A.} did not even mention Rossi's experiments, which were in fact the main target of his criticism, as Rossi, since the beginning of his activity in cosmic-rays researches, had been the principal proponent of the corpuscular hypothesis.  As Rossi later recalled, after  the nuclear physics conference held in Rome in 1931, Millikan\index{Millikan, Robert A.} ``refused to recognize'' his existence \cite[p. 18]{Rossi:1990aa}.

It is to be recalled that during those same years Millikan asked his student Carl Anderson  \cite{Anderson:1982ys}  to  measure the energies of the secondary charged particles produced by cosmic rays, in order to support his theory of primary cosmic rays as super-$\gamma$ radiation of well definite energies, which he had formulated since the end of 1920s starting from the interpretation of his own data \cite{Millikan:1928aa} \cite{Millikan:1928ab} \cite{Millikan:1928kx} \cite{Millikan:1928lr} \cite{Millikan:1928vn} \cite{Millikan:1930fk}. Few years would elapse before Anderson discovered the positron in 1932 \cite{Anderson:1932fk}, with no real knowledge of the new quantum mechanics and no awareness of its connection with the relativistic theory of the electron proposed by Dirac\index{Dirac, Paul A. M.}  since 1928 \cite{Dirac:1928kx} \cite{Dirac:1929vn}.\footnote{It is to be remarked that Dirac himself recognized the existence of the anti-electron as necessary for his  theory only  when the mathematician Hermann Weyl\index{Weyl, Hermann} ---the latter also was convinced at first that the {\it holes} in the {\it Dirac sea} must be  protons--- had demonstrated with symmetry arguments that the anti-particles' mass must be identical with the electron's \cite[pp. 85--92]{Bonolis:2004sf}.}

On his side Rossi had attended the quantum mechanics course held in Florence by Enrico Persico\index{Persico, Enrico}, and together with the young theorist Giulio Racah\index{Racah, Giulio} compiled the notes for the same course. Dirac's work had just appeared when Giovanni Gentile, Jr.\index{Gentile Jr., Giovanni} and Ettore Majorana\index{Majorana, Ettore} in Rome immediately applied it in their investigations on atomic spectra \cite{Gentile:1928uq} and Edoardo Amaldi\index{Amaldi, Edoardo} used it for a work on the Raman effect \cite{Amaldi:1929vn}. Racah\index{Racah, Giulio} himself applied Dirac's\index{Dirac, Paul A. M.} theory to phenomena of interference \cite{Racah:1937kx}. This provides some examples of the extent to which the Italians were aware and abreast of  the latest theoretical developments as shown by the well known series of lectures on quantum electrodynamics held by Fermi\index{Fermi, Enrico}  since 1928. Reflections on these issues found an elegant formulation in his well known paper {\it Quantum Theory of Radiation} \cite{Fermi:1933kx} and were also at the core of his $\beta$-decay theory  \cite{Fermi:1934qy}. Ideas circulated quickly within the small community formed by the few Italian physicists involved in researches on the new physics. Rossi was thus ready to enter and appreciate the discussion with Hans Bethe\index{Bethe, Hans A.} ---then a young and brilliant theoretician--- whom he invited to held seminars in Florence while the latter was spending some time with  Fermi\index{Fermi, Enrico} in Rome. At the time Bethe played a crucial role in the understanding of interactions between electromagnetic radiation and matter,  and like Homi Bhabha followed with great interest Rossi's work.\footnote{An in-depth discussion of Rossi's experiments appeared in H. Bhabha's article on the absorption of cosmic rays of 1933  \cite{Bhabha:1933fj}.} His  seminal work  attracted in particular the attention of Werner Heisenberg\index{Heisenberg, Werner}  at the time deeply involved in the nascent elementary particle physics and in quantum field theory\index{quantum field theory} \cite[pp. 886--890]{Rechenberg:2010kx}.\footnote{Rossi's early role in the scientific community of the 1930s, which show his strong relationship with theoretical physicists of the time, is elucidated also in connection with a series of letters exchanged with Werner Heisenberg\index{Heisenberg, Werner} in 1931--1932,  \cite{Gembillo:1993lr}.} The latter was deeply interested in cosmic rays and discussed Ross's early contributions in his well known 1932 article  \cite{Heisenberg:1932ys}.
Notwithstanding Millikan's\index{Millikan, Robert A.} critical views on Rossi's results, the close link with the paramount figures of the time throws light on how Rossi and his work were highly considered by theoreticians who were trying to frame  that mysterious phenomenology into the new born field theory. Some formidable problems had to be tackled,  which would eventually find a solution many years later.

The investigation of cosmic rays was facilitated by the fast coincidence method of multiple counter discharges developed by Rossi and by its application to the counter-controlled expansion of a cloud chamber by Patrick M. S. Blackett and Rossi's student Giuseppe Occhialini,  a perfect fusion of  ``image'' and ``logic'' \cite{Blackett:1932mz}. These two methods were particularly useful for investigating the interaction of charged particles with matter, because an impinging charged particle is selected by the discharges of aligned counters in coincidence. Blackett\index{Blackett,  Patrick M. S.} and Occhialini\index{Occhialini, Giuseppe Stanislao} were able to ``see''  the groups of particles whose existence Rossi had clearly demonstrated in his experiments as  some sort of a ``rain'' in the cloud-chamber, hence they named them ``showers''. In their second paper \cite{Blackett:1933fj}, Blackett and Occhialini pointed out that the occurrence of these tracks was a well known feature of cosmic radiation and was ``clearly related to the various secondary processes occurring when penetrating radiation passes through matter.'' They also credited Rossi for having been the first to investigate these secondary particles using counters. Blackett and Occhialini were able to confirm the interpretation proposed by Anderson \cite{Anderson:1932fk}  (``it seems necessary to call upon a positively charged particle having a mass comparable with that of an electron''), but they also placed the positive electron ---the ``Dirac hole-theory particle''--- into theoretical perspective. Basing on the clear evidence of the process known as ``pair creation'' provided by their counter-controlled chamber they were able to confirm Dirac's relativistic theory of the electron and were the first to expound the pair formation mechanism derived from experiments, which was the process underlying the formation of electromagnetic showers, one of the most striking facts of the phenomenology related to cosmic rays.

This kind of showers found its place in the theory at the end of 1930s when it was cleared that the soft component of cosmic radiation is constituted of a cascade of electrons and photons. In fact, after the discovery of the positron ---and in particular of the phenomenon of couple formation--- theoreticians began to pay much attention to cosmic rays as important to the development of theory, and cosmic-ray experimenters recognized quantum mechanics as an important tool, closely related to their experiments.  
In the period when understanding was far from clear and quantum electrodynamics appeared to break down at the high energies involved in cosmic rays, experiments using counters became one of the fundamental tools for testing the new physics. A theory providing a natural and simple explanation on the basis of the quantum electrodynamics cross-sections calculated by Bethe and Heitler \cite{Bethe:1934fj}  appeared in early 1937 in simultaneous papers by J.\ Franklin Carlson and J.\ Robert 
Oppenheimer \cite{Carlson:1937kx}, and by Homi Bhabha and Walter Heitler \cite{Bhabha:1937yq}. The theory explaining shower formation and the identification of the hard component with a brand-new particle, the mesotron, resurrected relativistic quantum field theory toward the end of the 1930s \cite{Brown:1991fk}.

The logic tradition was instrumental in dealing 
with large samples of events connected to a single kind of phenomenon.  And indeed, Bahbha\index{Bhabha, Homi K.} and Heitler\index{Heitler, Walther H.} used the ``Rossi transition curve''  to establish the correctness of their theoretical results \cite{Bhabha:1937yq}. The core of the theory was the explanation of the activity regarding the soft part of cosmic rays, the unequivocal success of Quantum Electrodynamics, which nevertheless left unsolved the study of the hard component leaving it as a challenge for the future.

On the whole, the work Rossi conducted in his Italian period shows an uneven character. After a great effort which led to several remarkable achievements, he moved to Padua in the autumn of 1932 and that would change the course of his life. Teaching responsibilities, and above all the burden of being in charge with the designing and construction of the new Institute of Physics, consumed his energies, even if during this period he carried on his expedition in Eritrea with Sergio De Benedetti\index{De Benedetti, Sergio}, during which  they measured the East-West effect\index{East-West effect}  Rossi had predicted in 1930 \cite{Rossi:1934aj} \cite{Rossi:1934rt}. Unfortunately,  he was beaten by Arthur H. Compton\index{Compton, Arthur H.} and Louis W. Alvarez\index{Alvarez, Luis W.} \cite{Alvarez:1933yq}, and by Thomas Johnson\index{Johnson, Thomas} \cite{Johnson:1933fj}, who were more lucky in getting sooner their research funds to perform this test which became a crucial one in beginning to unravel  the nature of cosmic rays. The study of the asymmetry predicted by Rossi confirmed the corpuscular nature of the primary radiation and revealed that the primaries, important in producing effects near sea level, were mostly positively charged.

As a by product of his experiments in Eritrea, Rossi observed what would be later called Extensive Air Showers\index{Extensive Air Showers}. He is not credited with this discovery, because the phenomenon was later re-discovered and deeply studied by Pierre Auger, who is officially considered the father of this kind of researches \cite{Auger:1938rt} \cite{Auger:1939aa}.

%CHIEDERE A GEORGE REFERENZA NUOVA BIO NEL BIOGRAFICO

During those years, Rossi established an intense relation with Enrico Fermi\index{Fermi, Enrico}   who appreciated the innovative character of Rossi's research and supported him to win the competition for the chair of experimental physics in the spring 1932. Their dialogue started since the very beginning of Rossi's scientific activity,  and emerged in different crucial occasions  both in the field of scientific activities and from a human point of view, as well as in connection with dramatic events in the history of the past century. Their very special fellowship interrupted in 1954,  when Fermi\index{Fermi, Enrico}  passed away. There was a red line between the two since the beginning because Fermi\index{Fermi, Enrico}   strongly appreciated Rossi's pioneering work and was constantly interested in  cosmic rays until his premature death in 1954. 

%an aspect that has been neglected so far from an historical point of view. The scientific community has naturally always dedicated a great attention to Fermi's\index{Fermi, Enrico}  articles on the origin of cosmic rays, among the most cited works in the scientific literature \cite{Fermi:1949dq} \cite{Fermi:1954aa}.

As stressed by Gilberto Bernardini\index{Bernardini, Gilberto}, Rossi's former collaborator in Arcetri, the natural development of physics in Italy after World War II, at least in some branches at the forefront of research, is not to be directly related with the works and teaching of the great scientific personalities of the past Italian tradition: ``More than of a tradition one should talk of schools. As far as Physics is concerned, two were the schools which had a crucial influence on the present: the one which was born in Rome around Enrico Fermi's\index{Fermi, Enrico}  extraordinary personality and intelligence; and the other one, which was born in Florence, and for which Bruno Rossi, great scientist and master, is essentially to be credited.'' \cite[p. 1]{Bernardini:1962qy}.

In analyzing Rossi's figure on the whole, it is to be emphasized, too, that in planning his research activity at the beginning of the 1930s, Rossi was preparing to assume a role and an identity within the scientific community which he would not abandon up to the 1950s, until the cosmic rays constituted a privileged source for high-energy particles, prior to their substitution with the beams artificially produced in accelerators.  By the early 1950s Rossi had become an authority in particle physics: adviser at the National Brookhaven Laboratories ---among the first to build  big accelerating machines between 1940 and 1950--- he was among the few experimental physicists invited to attend the first theoretical physics meetings with the major figures at the time, notably in Shelter Island, in 1947. Obviously, he played a prime role in later congresses of historical importance, as Bagn\`ere de Bigorre. For over twenty years Bruno Rossi gave a great contribution to the understanding of elementary particles, and above all he was also the distinguished builder of experimental devices as well as an elegant experimenter. And yet, those first years in Arcetri remained his true ``magic'' moments.

\section{New physics in a new world}

Supported by Fermi, Rossi had won a competition for the chair of Experimental Physics and worked in Padua from 1932 to 1938. At this time, after having just completed his work overseeing the design and construction of the new Physics Institute, he was rewarded by being denied the right to work there because he was a Jew. When the anti-semitic laws enacted by the fascist regime of Benito Mussolini passed in Italy, Rossi was dismissed and had to emigrate from his country with his young bride Nora Lombroso.\footnote{Unpublished correspondence and documents related to Rossi's emigration from Italy have recently been donated by the family to the MIT Archives. This material has thrown new light on those dramatic months preceding his decision to remain in U.S. after his short stay in Copenhagen and Manchester  and on the beginning of his new life on the American Continent \cite{Bonolis:2011fk}.}  After a short but scientifically productive stay in England with Blackett\index{Blackett,  Patrick M. S.},  Rossi moved to the U.S. on an invitation of Arhur Compton\index{Compton, Arthur H.}. Fermi\index{Fermi, Enrico} and Rossi's emigration from Italy in 1938 marks a watershed in the history hereby presented. They were bearing a heavy human burden and at the same time were aware of the innovative character of their researches which had put Italy  at  the forefront of world physics.
As Fermi\index{Fermi, Enrico}  landed in the States in early January of 1939, he had just been awarded the Nobel Prize for his work on artificial radioactivity induced by neutrons. Yet, he was experiencing some personal and unconfessed struggle. Despite being the world best expert in neutrons, he had missed to understand the existence of the just discovered phenomenon of fission which  the German chemists Otto Hahn\index{Hahn, Otto} and Fritz Strassmann\index{Strassmann, Fritz} had announced during the very first days of  his arrival in New York. Fermi was disappointment also because he was used to be ``the first of the class'' since when he was a boy.  Rossi, too, felt he had been in presence of ``big things'', albeit he had not been able to say he had made a definite ``discovery''.  It was not a chance that, as they landed in the American territory, they both engaged ---``furiously'' one would say--- in remarkable researches which would constitute noteworthy chapters in the history of physics. Fermi's\index{Fermi, Enrico}  work ended up with the first controlled chain reaction in December 1942, while Rossi in the period 1939--1942 conducted a series of experiments with different collaborators on the decay of the first unstable particle discovered in cosmic rays in 1936--1937, the $\mu$ meson, or ``mesotron''  as this particle was then called. These experiments  produced the final proof of the radioactive instability of mesotrons, established a precise value for its mean life, and as a by-product, verified for the very first time, the time dilation of {\it moving clocks} predicted by Einstein\index{Einstein, Albert} relativity theory \cite{Rossi:1940gf} \cite{Rossi:1941wd} \cite{Rossi:1942ai} \cite{Rossi:1942dp}. By 1942 the evidence for the decay of the mesotron at the end of their range had changed from the first two cloud-chamber tracks photographed by Williams and Roberts in 1940 \cite{Williams:1940rt} to the curve presented by Rossi and Nereson, which contained thousands of decay events and showed an exponential decay with a lifetime of about 2 $\mu$s  \cite{Nereson:1943bh}. The style and elegance of these achievements have been unanimously recognized  in the history of experimental physics \cite{Bonolis:2011fk}. These experiments, completed by Rossi during the war, symbolically closed an era that he himself called ``the age of innocence of the experimental physics of elementary particles'' \cite[p. 204]{Rossi:1983uq}. It was still a time when fundamental results for elementary particle physics could be obtained through extraordinarily simple experiments, which cost a few thousand dollars, and which required the help of one or two young graduates.

These experiments had their counterpart in the investigations on the fate  of mesotrons coming to rest in matter, which during the war years were carried out in 
Rome by Marcello Conversi, Ettore Pancini, and Oreste Piccioni. The Rome experiment was initially inspired by Rasetti's approach using the delayed 
coincidence method to signal the decays of stopped particles \cite{Rasetti:1941ys}, with one important variant, the introduction of the magnetic-lenses array already 
used by Rossi in 1931 \cite{Rossi:1930ve}. This  crucial experiment, conducted in the best logic tradition style, showed that the mesotron interacted with nuclei $10^{12}$ times more weakly than expected and could not possibly fulfill the role of the Yukawa particle, thus providing the first hints 
of a much more complex underlying reality. Luis Alvarez\index{Alvarez, Luis W.} mentioned this achievement in his Nobel Lecture as the very origin of modern elementary particle physics.\footnote{``I would suggest that modern particle physics started in the last days of World War II, when a group of young Italians, Conversi, Pancini, and Piccioni, who were hiding from the German occupying forces, initiated a remarkable experiment. In 1946, they showed that the `mesotron' which had been discovered in 1937 by Neddermeyer and Anderson and by Street and Stevenson, was not the particle predicted by Yukawa as the mediator of nuclear forces, but was instead almost completely non-reactive in a nuclear sense.'' http://www.nobelprize.org/nobel\_prizes/physics/laureates/1968/alvarez-lecture.pdf.} 

The riddle was definitively solved in 1947, 
when Cecil Powell, Cesare Lattes, and Occhialini at the University of Bristol 
identified a new particle in photographic emulsions exposed at high altitude. It became clear that the just discovered $\pi$-meson was the real Yukawa meson, 
and that the mesotron ---later dubbed the $\mu$-meson--- was the product of its decay \cite{Lattes:1947kx} \cite{Lattes:1947ve}. Indeed, as Bruno Pontecorvo\index{Pontecorvo, Bruno} was the first to understand, the observed behavior of this new particle was the first hint of  the existence of what would be later classified as the lepton family including the electron and the muon \cite{Pontecorvo:1947fk}. Such unexpected discovery definitely marked the beginning of a new phase in the field of elementary particle physics.

Rossi's work within the American scientific community starting at the end of the 1930s is also striking as for the change in his research style, since he started to systematically work with young collaborators who would become a new generation  of physicists. In a short time, many of them would gain  further new skills during  war research. In so doing Rossi, as well as Fermi\index{Fermi, Enrico}  in a different context, began to leave behind him a ``trail'' of disciples ---often colleagues-to-be--- who, as Canizares\index{Canizares, Claude R.} put it, would share his same spirit, together with a high  sense of scientific integrity \cite{Canizares:2006aa}: ``In an age of flamboyance and stridency,  Bruno's quiet character held sway. He proved time and time again that the understatement of true quality triumphs over exaggeration and self-promotion.'' That is a noteworthy testimony in terms of his teaching style ---human and intellectual--- shaping his figure of scientist.

The years from 1939 to 1943 form a self-contained period of Rossi's personal life and of his scientific activity, because it began with his arrival in the United States, as an exile from fascism, and ended with his shifting from the peace-time work at Cornell University to the war-time work at Los Alamos. It was almost all occupied by a research program on the spontaneous decay of mesotrons.\footnote{ When Rossi began to be active, world physics was still mainly European and German was the main scientific language, so that  it is to be remarked that during the ``Italian period'', Rossi wrote his papers in Italian, English and German, the latter being a language  widespread in Italy during the Fascist era. The beginning of a new phase in Rossi's life is marked by the systematic publication of scientific papers written in English, and generally appearing on \textit{Physical Review}. About 20 articles are related to this period, most of which are dedicated to the problem of the mesotron decay \cite{BlackettRossi:1938uq}  \cite{Rossi:1939fk}  \cite{Rossi:1939lr}  \cite{BlackettRossi:1939kx}  \cite{Rossi:1940fk}  \cite{Rossi:1940gf}  \cite{Rossi:1940pd}  \cite{Janossy:1940rt}  \cite{Rossi:1940mz}  \cite{Rossi:1940rr}  \cite{Rossi:1940zr}  \cite{Rossi:1941wd}  \cite{Rossi:1941la}  \cite{Rossi:1941la}  \cite{Rossi:1942eu}  \cite{Rossi:1942qe}  \cite{Rossi:1942ai}  \cite{Rossi:1942dp}  \cite{Rossi:1942eu}  \cite{Rossi:1942qe}  \cite{Nereson:1943bh}.}

 The hastening of events and the United States entering the war saw the Italian physicists responsible in being among the makers of the first nuclear arms ever built in history. The experimental work Rossi conducted during the war years saw also his participation in the research on radar, a chapter he barely mentioned in his autobiography. as a part-time consultant on radar instrumentation\index{radar}  at the Radiation Laboratory\index{Radiation Laboratory|see{MIT}}\index{MIT!Radiation Laboratory} of the {\it Massachusetts Institute of Technology} (MIT)\index{MIT}. From the late spring of 1943, he participated to the Manhattan Project\index{Manhattan Project} at Los Alamos\index{Los Alamos}  where the best brains were gathered to work on the production of nuclear weapons. With Hans Staub\index{Staub, Hans} he was responsible for development of detectors for nuclear experiments. The \textit{fast ionization chamber}\index{fast ionization chamber} was used in a series of tests conducted by Rossi that validated the implosion method\index{implosion method} for detonating the plutonium bomb\index{plutonium bomb}  and in a measurement he made of the exponential growth of the chain reaction in the Trinity Test\index{Trinity Test}  the first nuclear explosion, on July 16, 1945.
 
 Despite his very disapproving reaction to the initial bombing of the Japanese cities and his strong desire to leave Los Alamos, years later (in the 1980s) during the active historical research on that period, Rossi would visit the archives at Los Alamos and dedicated his time consulting various documents, some of them not even completely declassified, and wrote an article for the {\it Scientific American} recalling his participation in the Manhattan project. Thus, a new sensibility and awareness developed in relation to the high quality of that work and the willingness and capacity to respond to challenges in utterly new problems, in keeping the pace with time. A situation which, however, had left the physicists with a dramatic ``loss of innocence''. 

The importance Rossi had acquired within the American scientific context can also be measured by the offer of the Massachusetts Institute of Technology at the end of the war. After completing a book on ionization chambers, coauthored by Hans Staub\index{Staub, Hans} \cite{Rossi:1949fk},  Rossi left Los Alamos\index{Los Alamos} and moved to one of the most prestigious institution in the United States where he established the Cosmic-Ray Group  in the Laboratory for Nuclear Science and Engineerings to carry out research on the greatly increased scale made possible by the new availability of government support and the recent advances in technology. In these circumstances a new period of his life began, starting from 1946. Under Rossi's leadership, the Cosmic-Ray Group\index{MIT!Cosmic-Ray Group} carried out a wide range of investigations aimed at determining the properties of the primary cosmic rays, elucidating the processes involved in their propagation through the atmosphere, and measuring the unstable particles generated in the interactions of cosmic rays with matter. him to constitute and guide a research group to investigate cosmic rays in the newborn Laboratory for Nuclear Science and Engineering. The  great expansion program which characterized MIT as a natural prosecution of war activity, the prospect of ample funding, and particularly the possibility of a position for young people who had followed him from Los Alamos, were the main reasons for Rossi's choice of the Institute.\footnote{Interview of the author with Nora Rossi, July 26, 2007.} He certainly felt it was an opportunity  to catch up whatever he had left back in Padua. In accepting the Massachusetts Institute of Technology offer, he began a new life as a teacher and as a leader of the Cosmic-Ray Group he founded, as well as an inspirer of new generations of young physicists. Excited by the return to the active research and by the possibility to form a new generation, he  started to think on a grand  scale, relying on the  large economic resources which physicists had at their disposal after the scientific achievements of war research. 

The vast and complex program on cosmic rays and elementary particles saw him surrounded by an increasing number of young researchers, often coming from abroad to enrich the cultural circle within the local scientific community. Rossi was very active and enthusiast in promoting and guiding, though he was not directly involved in the experimental research itself. Apparently he had turned into someone quite different from the young researcher who in Arcetri used to file data in his notebooks, day after day, and to  build his instruments on his own. It was time to acquire a different sense of research, his main task had become to guide and promote, directing experiments and suggesting solutions to the construction of experimental apparatuses. The main line of his scientific thought is entangled with  the scientific activity of his collaborators, and of other research groups connected with him, on the background of the special relationship between science and society that emerged from World War II.  He constantly followed his collaborators' work,  and promoted new activities, becoming one of the leading persons of a worldwide research network, as the extraordinary number of letters he exchanged during his whole scientific life, is largely testifying. 

As stressed by his collaborator Robert Thompson\index{Thompson, Robert W.} ``Once settled at M.I.T., Rossi built a school of cosmic-ray physics such as had never been seen before.''\footnote{R. Thompson\index{Thompson, Robert W.} in ``First round-table discussion'' \cite [p. 274]{Brown-Hoddeson:1983uq}.}
 The following lines give us an idea of Rossi's personal feelings at that time \cite [p. 101]{Rossi:1990aa}: 
 
\begin{quotation}
\small
I was well aware that in my new position, my activity would be very different from what it had been in past years. Then, working alone or, at most, with the help of a few students, I would build the instruments needed for my experiments, I would take them to the place where they had to be used, I would make the measurements, and analyze the results. Now I had the responsibility of an entire group, and what mattered was no longer my own work but the work of the group. In the first place my task was to identify the most promising research programs among those that were within our reach. I had then to help where help was needed in the planning of the instrumentation and evaluation of the experimental results, all of this without discouraging the individual initiative of the researchers

\end{quotation}
\normalsize

At the time cosmic rays were really on the forefront of  ``particle physics'', since low energy processes in nuclear physics did not appear to be a source of knowledge about the basic nuclear forces.  It was not known what was going on with the mesotron of the cosmic rays and the positron, discovered before the war.  Modern elementary particle physics as such, was on the verge of becoming an independent field, but still strongly depending on cosmic ray research, whose instruments consisted largely of cloud chambers, ionization chambers and nuclear emulsions. A part of this equipment was mounted on trucks and taken to stations at high altitude, or operated in balloons, and flown in airplanes provided by the Navy. 
The fundamental problem  to be answered by these experiments was that of the nature of the meson, and in particular the mechanism of its disintegration. The renewed interest in fundamental research after the end of the war, set the stage for resuming studies on cosmic rays, which for about three years, had remained practically dormant.  If active research had been necessarily neglected during the war, fundamental advances in electronics and detectors had issued from war research. Rossi had been one of the leading figures in the field, so that the renewed research activity in cosmic ray physics had a strong continuity with his contribution to the war effort. An impressive research program aimed to study the production of mesons,  their interaction with matter, and their decay process by all possible means, included high-altitude mountain experiments, airplane measurements and the study of giant cosmic ray showers of the atmosphere, which would become one of the  MIT Cosmic-Ray Group\index{MIT!Cosmic-Ray Group}'s strongest research fields. At the same time, the problem of the identity of primary cosmic rays was tackled by means of suitable ionization chambers and amplifiers, carried by balloons.

Cosmic ray research fitted in with both high-energy quantum theory and nuclear physics. Its first remarkable connection with theoretical physics had been results regarding showers of electrons and gamma rays, which were explained by quantum electrodynamics. Later work on the mesons' decay processes, and on the interaction of cosmic rays with matter with the production of unstable particles, electrons and photons, had been instrumental in advancing knowledge on nuclear structure and nuclear forces. Investigation on mixed showers proceeded at the time along three main lines: penetrating showers, bursts, and extensive showers. The results of this work demonstrated the intimate connection existing between the soft and hard components.

By 1950 the first generation of students, most of them following Rossi from Los Alamos, were to leave or turn into collaborators. Cosmic rays and particle physics  were still merged, albeit the theory had serious difficulty in classifying all the particles discovered as well as in understanding the modalities of their decays. During those years Rossi and his group were running with the hare and hunt with the hounds: on one side research on cosmic rays with big and sophisticated cloud chambers and analogous devices to use on high mountain; on the other, detectors to be used with the new accelerators were built ---and Rossi himself was the adviser for the Brookhaven high-energy laboratories. On his side  Fermi had alway continued to keep an eye on cosmic rays, and in 1949, when he published his first article on the origin of cosmic rays \cite{Fermi:1949dq}, he was beginning to prepare for the developments in high-energy physics which were starting to come out of Berkeley. In 1953 he would be one of the first experimenters to use the 1.5 BeV Cosmotron, producing some well known fundamental results  \cite{Anderson:1952fk}. Fermi\index{Fermi, Enrico}  and Rossi demonstrated their ability to join imagination and competence in such a way as to deeply change the style of theoretical speculation and experimental  physics.  

During the 1940s, up to the advent of high-energy accelerators at the beginning of the 1950s, cosmic rays continued to play a leading role as a source of high-energy particles. In particular, they provided the great amount of data needed to test relativistic field theories, which came from the rich phenomenology of high-energy interactions involving particle creation and annihilation in these processes. The ``image and logic'' instrumental traditions emerging from the early studies on the nature of cosmic rays continued to develop, each with its own practitioners and with different styles, sometimes competing, but often collaborating in devising new methods of particle detection that could provide evidence for the variety of the subatomic world. 

Yet that was also the era of ``little science,'' the 
same science characterizing the simple tabletop experiments performed by Thomson and Rutherford at the dawn of the 20th century. As Rossi noted later ``[\dots] results bearing on fundamental problems of elementary particle physics could be 
achieved by experiments of an almost childish simplicity, costing a few thousand dollars, 
requiring only the help of one or two graduate students'' \cite{Rossi:1983uq}.  A most remarkable change in the second half of twentieth century science was its growth in scale, scope, and cost. No wonder that in the 1980s, in an era of big science, and as a mature representative of the old generation, Rossi felt ``a lingering 
nostalgia'' for what he called the ``age of innocence of the physics of elementary 
particles.''

\section{Cosmic rays and accelerators: the beginning of a competition}

Prior to the construction of modern accelerators, high-energy particles could be observed only in cosmic rays. This explains why cosmic rays remained up to the mid-1950s a most important subject of high-energy physics. Since the discovery of the positron in 1932, many new particles were first reported to occur in cosmic rays. The latter were used to investigate what happens when high-energy particles strike an atom or an atomic nucleus.

Up to the middle of the 1950s the ``little science'' program carried out by the cosmic ray physicists with cloud chambers and nuclear emulsions, continued to be competitive with ---and even to dominate--- the ``big science'' of the accelerators. But since some time Rossi's interest in the primary cosmic radiation had been aroused by experiments carried out toward the end of the 1940s. 
By the end of 1940s and beginning of 1950s, the secondary processes occurring in the atmosphere and generating the various components of the local radiation (penetrating component, shower-producing component, nuclear-active component) had been clarified, but cosmic-ray research was undergoing a radical change. On the one hand, in 1940 balloon experiments by Schein and co-workers \cite {Schein:1941zr}    using complex Geiger-M\"uller counter arrangements, and in 1948 balloon experiments using nuclear emulsions \cite {Bradt:1948aa}  \cite {Bradt:1948fk},  had solved the problem concerning the nature of the primary cosmic radiation, three decades after Hess' seminal discovery. It was thus found that cosmic rays contain nuclei of various elements including iron, whose energies amounted to many GeV. Protons make up about 90\% of the primary cosmic rays while helium nuclei are nearly 10\%, and all other nuclei to around 1\%.  Research on the primaries  began to stir up questions connected with the origin of cosmic rays, but at the moment problems related to nuclear interactions  were still overwhelming. In fact, as recalled by Rossi, the discovery of the strange particles and the attempts ``to establish their nature and behavior became one of the most fascinating and demanding activities of our group'' \cite [p. 117]{Rossi:1990aa}.

At that time the activities of the MIT Group were in fact divided between experiments on cosmic rays and experiments with accelerators. At the beginning of the 1950s a semblance of order began to appear, but  a much deeper understanding of nuclear forces still needed a lot of ingenuity both by theorists, and by accelerator experimenters and cosmic ray physicists. The five years from 1947 to 1952 were very special and ``unique in several respects'', as remarked by Robert Marshak\index{Marshak, Robert E.}: ``The cosmic-ray experimentalists, the theoretical particle physicists, and the high-energy-accelerator experimentalists joined together, on fairly equal terms, to gain an understanding of the subnuclear structure of matter'' \cite [p. 398]{Marshak:1983uq}.

With the development of high-energy accelerators, cosmic ray experiments were  ceasing to be the only  ---or even the most profitable--- source of information about high-energy interactions and about the ``ephemeral particles produced in these interactions'' \cite [p. 81]{Rossi:1982aa}. 
On December 21, 1951, Fermi's pion-nucleon cross-sections for the scattering experiments carried out with the Chicago synchrocyclotron, had provided evidence of the existence of what was his last fundamental discovery, the so-called $N$* ---or ``3-3''--- pion-nucleon resonance \cite{Anderson:1952fk}, the confirmation of which came some time after his death. The 3-3 resonance, in focusing  physicists' attention on isospin and the concept of symmetry, was fundamental for the development of particle physics. As remarked by Abraham Pais\index{Kramers, Hendrik A.}: ``theorists had at least something to offer their experimental colleagues'' \cite [p. 486]{Pais:1986fk}.   In bringing  to the fore the concept of symmetry, which would become a guiding principle for theoretical physics, these experiments, and their theoretical explanations, marked the beginning of a new era in the development of modern particle physics. 

 In this period, the MIT Cosmic Ray group and the \'Ecole Politecnique  group had a leading role in resolving the charged kaon group. Their cloud chamber were run at Echo Lake and at the Pic du Midi. This work was part of a series of ``excellent studies'' of high-energy cosmic-ray events in multiplate cloud chambers, in the course of which stopping kaons were found. By the time of Rochester III ( ``Annual Conference on high-energy Nuclear Physics''), held 18--20 December 1952, Marshak\index{Marshak, Robert E.} remarked: ``The machine results were certainly overtaking the cosmic ray results in pion physics and the theorists were frantically trying to develop a plausible theory of the pion-nucleon interaction starting with isospin invariance. But, absent the completion of the 3 GeV Cosmotron at Brookhaven, the cosmic ray physicists still dominated strange particle physics'' \cite {Marshak:1989aa}.\footnote{Rossi presided the Experimental Physics Session, whose results were both coming from cosmic-rays and accelerators.}
 
 Those years would culminate in the the 3rd International Cosmic Ray Conference, the great congress held at Bagn\`eres de Bigorre in 1953 which old people still remember with nostalgia. Not only was a major occasion of encounter for the groups involved in the study of elementary particle physics all over the world, but also was the occasion of examining the data collected in the previous years in an attempt to clarify a very confused situation in the field. Rossi's role was a towering one, according to many participants. Cosmic ray research and high-energy particle physics were still sharing a remarkable connection, even if the feeling that accelerators were beginning to invade the field began to spread during the Bagn\`{e}res-de-Bigorre Conference.  This watershed meeting marked the hour  point for physicists, like Rossi, who gradually had to face the choice between continuing to be a particle physicists working with beams of artificially accelerated particles, or choosing to work with cosmic rays which were more and more viewed as an astrophysical phenomenon. At the conference animated discussions and presentation of a great quantity of accumulated data contributed  to disentangle in part the complex reality revealed by the great variety of forms in which the new particles ---as well as the great variety of their end products---manifested. As recalled by Milla Baldo Ceolin, at the time a young researcher of Padua University, ``order began to emerge from chaos'' \cite [p. 7]{Baldo-Ceolin:2002aa}.  Milla Baldo Ceolin well remembered how Rossi dominated the scene at Bagn\`ere de Bigorre.\footnote{Interview with Milla Baldo Ceolin,  November 10, 2006.}
 
 The titles of international  Summer schools organized in Varenna\index{Varenna} in 1953 and 1954 well express the changing times and the fast evolution of research: from ``Questions related to the detection of elementary particles, particularly regarding cosmic radiation'', to ``Questions related to the detection of elementary particles and their interactions, particularly regarding artificially produced and accelerated particles''.\footnote{The collection of lectures held in Varenna ({\it Rendiconti del corso tenuto a {Varenna} nel 1954}) was published in the {\it Supplement} of {\it Il Nuovo Cimento}  series X, August 1955.}

 In October 1954 the G-stack, a giant stack of 250 sheets of emulsions which had a dimension of $37\times27$ cm$^{2}$ in order to follow decay product to the end of their ranges, flew for six hours at 27,000 meters of altitude.  The emulsions sheets were distributed between the participating laboratories: Bristol, Milano and Padova. Their results were announced at the Pisa conference in the summer of 1955.

The cosmic ray physicists could be proud; they had found just in time all possible decays of the heavy mesons, and made it very plausible that there was one and only one $K$ particle. Rossi's ``wisdom'' and ingenuity received a confirmation at the Pisa Conference in July 1955. The G-stack experiment, signed by 36 authors instead of the few typical of those times, was the last great experiment of cosmic rays in particle physics.\footnote{93 kg of emulsion, were flown from a balloon. The results were measured and analyzed by a collaboration formed by a large number  of laboratories, twenty-one European and one Australian.}
 Actually, it showed that the $K$-mesons, having different decay modes, had masses that were equal within a few tenths of 1\%. 

 Their triumph was a swan's song.  In the meantime the new generation of machines had begun to produce results comparable to those obtained by cosmic ray physicists. 
 In 1953 the first strange particles produced in laboratory experiments using accelerator beams were observed. 
 At the same Pisa conference the first results of the Berkeley Bevatron were announced \cite {Fowler:1953eu}  \cite {Fowler:1953fp}, which brought better proofs of Rossi's idea that the different decay modes were due to the same particle.
 
 It was just the beginning. From the mid-1950s, the interest in cosmic rays shown by high-energy physics fell sharply mainly because high-performance accelerators became available for experimental studies. In this atmosphere, cosmic ray research underwent an evolution which gradually changed its old character. Some scientists turned to work with accelerators. Others applied themselves to cosmic ray problems different from those that had been in the foreground in earlier years. Interest became increasingly focussed toward the astrophysical aspects of the cosmic-ray problem.

 The discovery of the antiproton in the middle of the 1950s is a significative example of  the transition from cosmic rays style of research, to the authoritative proofs provided by the more systematic study with accelerators. And not only. Power and funds more and more concentrated around big accelerating machines,  which  attracted the attention both in terms of money and prestige.  Because of the doubtful interpretation of the the so-called ``Faustina event'', a double star found in one of the emulsions exposed to cosmic rays during the G-Stack balloon flight,\footnote{According to  Alberto Bonetti, who at the time was asked for an opinion, it was difficult to give a definite interpretation and he gave the antiproton at 50\%. Interview of the author with Alberto Bonetti, April 10, 2007.}
Edoardo Amaldi\index{Amaldi, Edoardo} wrote to Emilio Segr\`e asking for a collaboration. Segr\`e was ``impressed'' by `Faustina',\footnote{E. Amaldi to E. Segr\`e, March 29, 1955; E. Segr\`e to E. Amaldi, April 15, 1955, Amaldi Archive, Physics Department, University  `La Sapienza', Rome, Box 455.}
but he also feared that the antiproton race might be won by cosmic ray physicists, and that his efforts in having a specific experiment at the Bevatron funded by the Atomic Energy Commission  might completely vanish. What can  be defined as the ``demonstration of the existence of the antiproton'' \cite {Chamberlain:1955ab},  became a glory of the Lawrence's Radiation Laboratory.\footnote{The whole story, and particularly the relationship between Segr\`e and Amaldi\index{Amaldi, Edoardo} on the occasion of this collaboration, has been insightfully reconstructed by Battimelli and Falciai \cite {Battimelli:1995yq}.}

The new unstable particles had been the central problem in the agenda of the Bagn\`eres-de-Bigorre conference,  the last conference  where  elementary particle physics data came entirely from cosmic ray work. In great contrast, this problem was not discussed at all during the 4th International Cosmic Ray Conference held in Guanjuato, Mexico, which was devoted to cosmological and geophysical problems. 
  As remarked by Peyrou, ``Two year later the cosmos took its revenge. The cosmic ray conference in Mexico city was entirely devoted to topics other than new particles and it is at another conference, at Pisa, that the cosmic rays celebrated a final triumph in that field, only to abandon it definitely to accelerators''  \cite [p. C8-29]{Peyrou:1982aa}.

The search for particles and their mysterious ways of decay would continue without them starting from the second part of the 1950s, when the interest in the fascinating world of the microscopic structure of the matter became to be a research field for accelerator physicists.

 In June 1952 Rossi had completed a most arduous task of preparing the volume \textit{High-Energy Particles} \cite {Rossi:1952aa}, a full survey of the information gathered up to 1951 about the elementary particles.\footnote{In what was an advanced review textbook, as well as a precious source of references, Rossi presented the most important experimental data, together with many theoretical developments. Over 500 references were quoted, two-thirds of which had been published since 1945.}  
 high-energy physics in 1952, as defined by Rossi in his book, dealt with particles of energy greater than 10 GeV. The study of such particles, since the beginning of the 1930s had been the exclusive domain of cosmic-ray physicists,  but with the development of high-energy accelerators  an enormous amount of data was provided by the new opportunity for more controlled and precise experiments. At the time, laboratory sources of particles with energies up to several hundred MeV were already at disposal of experimenters, and some of the machines under construction were going to reach energies of the order of GeV. ``However, remarked Rossi, most of our present knowledge concerning the properties of high-energy particles was obtained through the study of cosmic-ray phenomena and, for a long time to come, cosmic rays will probably retain their monopoly over the energy region beyond 10 Bev'' \cite [p. 6]{Rossi:1952aa}. Being interested in dealing with cosmic rays mainly as a source of high-energy particles, the phenomena of air showers were outside the scope of the subject, ``even though ---stressed Rossi in the concluding lines of his introduction--- the study of air showers is the only source of information on the properties of particles with energies of the order of $10^{15}$ ev or more.''

Already in 1949, in presenting the status of cosmic ray research,  Rossi well expressed the double nature of cosmic-ray physicists, which actually was a representation of his own scientific identity:\footnote{B. Rossi, ``Present status of Cosmic Rays'', draft of a talk, Rossi Papers, MIT Archives, Box 22, Folder ``1949''.}

\begin{quotation}
\small
[\ldots] one can look at cosmic rays as an {\it astronomer} or as a {\it physicist} [emphasis added].

If you look at cosmic rays as an astronomer, you are likely to ask yourself questions like: what and how are cosmic rays produced? How are they distributed throughout the universe? How large is their energy, confronted with other forms of energy in the universe? What is the influence on cosmic rays of the magnetic field of the galaxy, of the Sun, of that of the Earth? Can we learn anything about the structure of these fields from the study of cosmic rays?

If you look at cosmic rays with the eyes of the physicist, you mainly see in the cosmic radiation a source of very high-energy rays [\dots]   that you  can use to obtain information on the properties of elementary particles on the one hand and on the properties of matter, and more particularly of atomic nuclei, on the other hand [\dots]

\end{quotation}
\normalsize

\section{Outer space}

The life of Rossi's group at MIT developed at fast pace producing remarkable results, especially in the field of Extensive Air Showers which had been one of the main research lines explored by the group since the beginning. New and old techniques opened the way to new approaches able to determine the energy of primary cosmic rays which had been studied since the end of the war with the use of rockets, air balloons and aircrafts disposed by the Navy. At  the end of the 1940s remarkable results had provided hints of the existence of particles with an energy of about $10^{17}$ eV, showing that a  small number of fast particles from space were more energetic than the particles produced by the largest man-made accelerators. They were revealed by the showers of other particles they create in the air. The most important result of air shower experiments was in fact the information they provided about the primary very high-energy particles by which they are produced.

 These rare events attracted the attention on the origin of such high energies. This central issue was related to a series of questions such as the sources of production, the way of propagation and the transformations cosmic rays undergo in the interstellar medium, and the mechanism of their acceleration. The transition to a new stage of Rossi's scientific life is testified by his keen interest in the origin of cosmic rays.  During the second Varenna course of 1954 Rossi held two lectures---``Fundamental Particles'' and  ``Origin of Cosmic Rays'' \cite{Rossi:1955fk}  \cite{Rossi:1955lr}--- whose titles testify the ongoing evolution of his identity as a cosmic-ray physicist.  The origin of cosmic rays was becoming a hot problem also because, on the other hand, cosmic ray physicists studied the composition, the energy spectrum, the time variations and the directions of arrival of primary cosmic rays in an effort to get some clues on their place of origin, on the mechanism responsible for their production, and on the conditions prevailing in the space through which they travel on their way to the earth. A field in which the Cosmic-Ray Group\index{MIT!Cosmic-Ray Group} had been very active, in parallel with research in the field of atomic and subatomic physics.

The discovery of the existence of particles characterized by energies much  higher than could be previously imagined, revived the interest for mechanisms able to accelerate cosmic rays at such energies. Back in 1949 Fermi\index{Fermi, Enrico}  had published a most cited article  \cite{Fermi:1949dq} where he proposed for the first time  a model for the acceleration mechanism responsible for the production of cosmic rays which could be compared with data. At that time, attention began to switch to the outer space and astrophysical aspects started shaping the new path for cosmic rays. Between 1953 and 1954 the origin of cosmic rays had become a hot topic, various theories were published. Fermi\index{Fermi, Enrico}  got interested in magnetic galactic fields \cite{Fermi:1953aa} and updated his hypothesis at the light of the new data \cite{Fermi:1954aa}. At the time Rossi, Philip Morrison and Stanislaw  Olbert were writing an article on the origin of cosmic rays \cite{Rossi:1954oe}, so that the issue was discussed in a correspondence  which was forcibly interrupted with Fermi's\index{Fermi, Enrico}  premature death in the late autumn of 1954. 

Magnetic fields, particles generated in the explosion of supernovae, and nuclei of elements expelled by astrophysical object of unknown nature were more and more crowding  the intergalactic space. In entering the field of high-energy astrophysics,  cosmic ray physicists were thus gradually  forced to change their identity.

During the 1950s some members of Rossi's cosmic rays group at MIT launched the production of new detectors like the plastic scintillators to investigate the big air showers generated in the atmosphere by primary cosmic rays. Such pioneering experimental methods  led to cosmic-ray research in the region of highest energies. By 1956 Rossi's group was measuring cosmic-ray air showers with an array of 20 large scintillation counters at the Agassiz Station\index{Agassiz Station} of the Harvard College Observatory. The area of the array was about a sixth of a square kilometer. The discovery of primary cosmic rays with energy as great as 5$\times$10$^{18}$ electron volts recorded in 1957 and arriving isotropically created a stir in the astrophysical community because of its bearing on the problem of cosmic-ray origin \cite{Rossi:1957xq}. Such phenomenon suggested that the highest energy cosmic rays originate outside of our galaxy. The setting up of a great array of plastic scintillator detectors  allowed the recording of a shower generated by a primary particle whose energy was estimated to be close to $10^{19}$ eV. With this event, the MIT Group was extending the energy spectrum of primary cosmic rays.  Combined with the data at lower energies collected by other groups, it allowed a series of considerations of astrophysical character.  Results from such large air shower experiments were presented in preliminary form on {\it Nature} \cite {Clark:1957fj} \cite{Clark:1957rt}.\footnote{For a reconstruction of MIT Air shower program see \cite{Clark:1985aa}.}  The evidence had showed no sign of a falling-off in the shower size spectrum beyond  $10^{8}$ particles. It was estimated that only one particle above $10^{16}$ eV was arriving per m$^{2}$ per year, so that the extension of the primary cosmic ray energy above $10^{18}$ eV, pushed Rossi and his group at MIT to design a much larger array to cope with the decreasing event intensity with energy.  The ``Volcano Ranch Air Shower Experiment'',  an array of 19 plastic scintillators arranged in a regular hexagonal pattern, whose diameter was initially 1.8 km (area 2.2 km$^{2}$) and was later increased to 3.6 km, at the beginning of 1960s provided evidence showing that the primary spectrum  extended to about 10$^{20}$ eV  \cite {Linsley:1962yq}. This result was to be considered a definite proof for the extra-galactic origin of some cosmic rays.

The years between the end of the war and the first 1960s see Bruno Rossi at the core of a worldwide network spreading out from MIT to India, South America, Japan and Europe, and it essentially presents a natural testimony of the entire activity of the cosmic rays group at the MIT, and later of the birth of the {\it Center for Space Research} which  constituted its natural heritage. Nonetheless, the passage to the second half of 1950s represents a crucial ---and perhaps even critical--- period for Bruno Rossi and his scientific activity. He even thought of dedicating his efforts to biophysics,  attracted by the fascinating field of molecular biology, as it was the case for many physicists between the end of the war and the first 1950s. According to his wife's  testimony, Nora Lombroso, he tried to approach the subject during a whole  summer and soon realized he could not achieve the excellency he always aspired to. 

Unexpectedly, the opening of the extraordinary technological window by the space era, officially inaugurated in October 1957 with the launch in orbit of the Russian Sputnik, in particular the discovery of the Van Allen belts, generated in Rossi as enthusiasm as he used to have during the {\it Arcetri era}. Indeed, it is not a chance that the first discoveries at the beginning of the space age were made by cosmic ray physicists. As technological progress opened new windows, Rossi would draw what he called {\it a subconscious perception of the immense richness of nature, a richness which goes beyond imagination}. Thus, that sensation generated his strong willingness and impetus in exploring nature with new eyes, with the hope {\it to see something unexpected}. 

In the meantime, his  {\it status} made him one of the most influential scientists in the United States, and saw him involved also in the U. S. science policy. The many documents related to those years depict his active role in the scientific institutions, and especially  in connection with the contemporary foundation of the NASA\index{NASA}, the American Space Agency. Traces of his importance in the scientific community, and not only that, can be found in the correspondence with the several protagonists of American space research; in particular, the correspondence with John Fitzgerald Kennedy\index{Kennedy, John F.} when he was getting ready for the election campaign which would end with his election as the 35th President of the United States of America in November 1960.

\section{From earth laboratories to physics in space}

As one of the fathers of modern physics, Rossi witnessed  the birth of particle physics, and was a pioneer as well as one of the main investigators in this fields for many years. He was also a pioneer in virtually every aspect of what is today called high-energy astrophysics. The evolution of cosmic-ray discipline led in fact to the genesis of two new sciences, high-energy elementary-particle physics and cosmic-ray astrophysics. While cosmic ray physics moved more and more towards astrophysics, Rossi continued to be one of the inspirers of this line of research. When outer space became a reality, he did not hesitate to leap into a new scientific dimension, which soon revealed its extraordinary richness. The realization of his ideas opened completely new perspectives and fields of research. 

%STORIA DI UN FISICO SPERIMENTALE, LA TEORIA E' STATA SPESSO LA CHIAVE DI LETTURA DELL'EVOLUZIONE DELLA FISICA, LIBRI DI GALISON SUL RUOLO DEGLI ESPERIMENTI
%UNITA' ED EVOLUZIONE NEL PERCORSO DI ROSSI (STYLE OF REASONING, VEDI SCHWEBERDARRIGOL QED (IN CARTELLA QED)

The main path leading Rossi and some of his collaborators from cosmic ray research related to particle physics to physics  in space goes through the 1950s up to about 1957, when accelerators had definitely  taken the power marking  a new era in the relationship between cosmic ray physics and high-energy particle physics. In following the path of giant experiments aiming at detecting the air showers produced by very high-energy particles coming from outer space, Rossi's and his collaborators' scientific interests changed during the years, and got more and more oriented toward astrophysical problems. Besides his advising and supporting role in the crazy activity following the Sputnik shock, Rossi began to think of the outer space in practical terms, thus promoting the design of new research instruments to study  physical entities, whose existence was considered doubtful, and  in any case to be first demonstrated; in fact, experiments he planned in order to put into practice his ideas  could be considered by some as not plausible. This visionary attitude has often been at the base of the success  of many enterprises, otherwise not even taken in consideration.

In the spring of 1958, as a response to Sputnick, the United States created the National Aeronautics and Space Administration (NASA)\index{NASA} and at the same time, the \textit{National Academy of Sciences} created a \textit{Space Science Board} (SSB)\index{Space Science Board} to interest scientists in space research and to advise NASA and the other federal agencies the \textit{Academy} expected to be engaged in space research. Bruno Rossi was among the fifteen members of the \textit{Board} called together for their first meeting in New York on June 27, 1958. 

 By this time it turned out to be very natural for the Cosmic-Ray Group\index{MIT!Cosmic-Ray Group} to redirect part of their activities toward the new field of science in space, when opportunities for this kind of experiments became available. However, it is to be remarked, that in Rossi's and his collaborators' opinion space vehicles were not to be used for the purpose of obtaining better information on primary cosmic rays than it was possible to do from earth-bound stations. Rather, they had clear that space experiments had to be devised that would provide a \textit{new} line of attack to the same astrophysical problems in which they had become interested through cosmic-ray studies. This led into two research programs, both related to cosmic-ray research proper: interplanetary plasma and $\gamma$-ray astronomy.

Bruno Rossi did not directly participate to the MIT satellite $\gamma$-ray astronomy\index{gamma-ray astronomy}  which was initiated by William L. Kraushaar\index{Kraushaar, William L.} in 1958, but he immediately promoted the project as a member of the Space Board Committee. George W. Clark\index{Clark, George W.} joined W. L. Kraushaar\index{Kraushaar, William L.}, and both designed and built a $\gamma$-ray telescope at MIT's Laboratory for Nuclear Science\index{MIT!Laboratory for Nuclear Science and Engineering} and directed the experiment which led to the first observations of high-energy ($>$ 50 MeV) cosmic rays  \cite{Kraushaar:1962qa}  \cite{Clark:1962qy}  \cite{Clark:1965uq}.  The appearance of $\gamma$-ray astronomy\index{gamma-ray astronomy} significantly broadened the possibility to study cosmic rays in the Universe, for cosmic-$\gamma$ photons with a sufficient high energy are generated only by cosmic rays especially as a result of particular processes. $\gamma$-ray astronomy\index{gamma-ray astronomy} could give information not otherwise obtainable, about nuclear processes occurring in the Sun and in other celestial objects; about interactions between high-energy particles and interstellar matter in distant regions of our galaxy and extragalactic objects.  

At the beginning of the space age, observations from the Earth had already suggested that the space surrounding our planet is not entirely devoid of matter, as had been supposed in the past, but rather contains a dilute plasma,  an ionized gas that was considered presumably consisting of electrons and protons. The Cosmic-Ray Group\index{MIT!Cosmic-Ray Group} at MIT had been concerned with this problem because of the possibility suggested by some scientists that certain temporal changes of cosmic-ray intensity might be due to clouds of magnetized plasma ejected by the Sun into the surrounding space. At the end of the 1950s most scientists were thus persuaded that a plasma of solar origin was filling interplanetary space all the way to the EarthÕs orbit and probably beyond.  However, the views concerning the properties of this plasma were widely divergent. Therefore, Rossi felt motivated to initiate a study of interplanetary plasma\index{plasma!interplanetary} with his own group at MIT. When he became a member of the \textit{Space Board Committee} he could immediately point out how no experiment had been devised to enquire about the conditions of space around our planet and of the Sun--Earth relation. The MIT Group planned an experiment to be performed in outer space by means of a plasma probe, a modified {\it Faraday Cup}, which was carried aloft by Explorer 10 on March 25, 1961. The \textit{probe} measured for the first time the speed and direction of the solar wind\index{solar wind} streaming past the Earth at supersonic speed, and yielded data that established the existence of a geomagnetic cavity, the magnetopause\index{magnetopause}  i.e. a region of space surrounding the Earth, which is shielded from the solar wind\index{solar wind} by the Earth's magnetic field. Under such circumstances the formation of a bow-wave was expected, pointing towards the direction of the oncoming wind and enveloping the geomagnetic cavity.  The final results were presented at the Third International Space Science Symposium held in Washington, D. C., from April 30 to May 9, 1962  \cite{Bridge:1960aa}  \cite{Rossi:1961aa}  \cite{Bridge:1962fj}  \cite{Rossi:1962kx}  \cite{Rossi:1963aa}  \cite{Bonetti:1963aa}  \cite{Bonetti:1963ab}  \cite{Rossi:1964aa}. 

The Mariner 2 mission launched on August 27 of that same year, carried a plasma probe built by the Jet Propulsion Laboratory which could measure the energy spectrum of both electrons and protons. It provided a great quantity of data showing that the solar wind is the normal state of things in interplanetary space and definitely confirmed Eugene Parker's prediction of the supersonic expansion of the solar atmosphere \cite{Parker:1958oq}.\footnote{Parker submitted his paper to the {\it Astrophysical Journal} in 1958, but it was rejected by two referees, and published thanks to the then editor Subrahmanyan Chandrasekhar.} The detector used in the first successful flight, the {\it Faraday Cup}\index{Faraday Cup}  continued to be used in later flights to reveal the properties of interplanetary plasma\index{plasma!interplanetary}  particularly around other planets of the solar systems. 
Under the leadership of Herbert Bridge, the MIT group later kept at the forefront of this exploration, one of the most active branches of space research.

In parallel with his interest on plasmas in space, Rossi Rossi became also convinced of the importance of exploring the X-ray window of the Universe. For centuries visible light from celestial objects has been the only source of astronomical information available to man. The only other portion of the spectrum capable of penetrating the atmosphere and the ionosphere was that corresponding to short-wave radio signals. The development of big radio telescopes in the 1940s made it possible to detect such signals, so that radio astronomy could enormously advance our knowledge of the Universe, particularly with the discovery of several classes of new objects, such as pulsars, quasars and radio galaxies, and thus became  able to  detect even the high-energy electrons produced thousands or millions of light years away from the Earth. 

Rossi's intuition, that a new window of the electromagnetic spectrum could, and should be opened thanks to space technology, inspired a program in X-ray astronomy\index{X-ray astronomy} at {\it American Science \& Engineering} (AS\&E)\index{American Science \& Engineering}  a company in Cambridge founded by Martin Annis\index{Annis, Martin},  Rossi's former student. At that time, Rossi was acting as a consultant for AS\&E as were, among others, George Clark\index{Clark, George W.} and Stan Olbert\index{Olbert, Stanislaw}. Riccardo Giacconi\index{Giacconi, Riccardo}  who had recently joined the Company, took charge of the program.  The challenge of developing an entirely new type of instrument was taken up, and a major effort aiming at the research of X-rays from celestial sources other than the Sun, was started in the fall of 1959. The effort to detect X-rays from celestial sources other than the Sun required completely new kind of detectors. A small group, initially formed by Rossi, Clark\index{Clark, George W.} and Riccardo Giacconi\index{Giacconi, Riccardo}  discussed the prospects for such a research and the necessity of achieving such greatly increased sensitivities in order to detect X-rays from such faint sources.\footnote{R. Giacconi, G. W. Clark, and B. Rossi, A Brief Review of Experimental and Theoretical Progress in X-Ray Astronomy, Technical Note of American Science \& Engineering, ASE-TN-49, January 15, 1960; R. Giacconi, G. W. Clark, Instrumentation for X-ray astronomy, ASE-TN-50, 15 January 1960; AS\&E proposal to NASA: Measurement of Soft X-Rays from a Rocket Platform Above 150 km, ASE-P-26, February 17, 1960.} They decided to follow two lines of attack. The first was the development of an X-ray telescope, even if it was immediately clear that such a novel instrument would require several years.\footnote{See patent 3,143,651 filed on February 23, 1961 by Rossi and Giacconi, and patented 4 August 1964: \textit{X--Ray reflection collimator adapted to focus X-Radiation directly on a detector} and \cite{Giacconi:1960aa}.} 
 The second line had the aim of improving the performance of the thin-window gas counters already used mainly by Herbert Friedman\index{Friedman, Herbert} for solar X-ray astronomy. Work on the improvement of thin-window X-ray counters, which extended over a period of about two years and was carried out to a great extent by Frank R. Paolini\index{Paolini, Frank R.},  produced a detector about 100 times more sensitive than the X-ray detectors used previously in solar astronomy. The group now included also Herbert Gursky\index{Gursky, Herbert}.\footnote{Proposed experiment for the measurement of soft X-rays from the moon, ASE-83-I, 25 October 1960; The feasibility of an assay of the lunar surface by means of soft lunar X-rays, ASE-89-I, 17 November 1960; R. Giacconi, Instrumentation for the detection of soft X-rays from the moon, ASE-96, 30 November 1960; Determination of rocket aspect and X-ray detector orientations, ASE-158, 29 August 1961.} 
Supported by the Air Force Cambridge Research Laboratories\index{Air Force Cambridge Research Laboratories} a rocket experiment  launched in June 1962  detected Scorpius-X1, an  unexpectedly bright X-ray source, some 9,000 light years away. Apart from the Sun, it revealed the most powerful X-ray source in Earth's skies, an object emitting a thousand times more X rays than the Sun at all wavelengths and a thousand times more energy in X rays than in visible light. The experiment also proved that the Universe contains background radiation of X-ray light.

An entire new astronomy was opening,  an essential source of information about new classes of stellar objects like neutron stars and black holes,  which has thrown light in previously unknown processes going on in the Universe. The discovery of the first extra-solar X-ray source, was presented August 1962 in occasion of the congress at Stanford University and was  welcomed with enthusiasm, though mixed with  ``considerable skepticism''. It appeared incredible that a very simple instrument could have revealed such an intense signal as to appear incompatible with the available astronomic data {\it regarding the celestial bodies then known}. It is worth noting that the article presenting those results was published in {\it Physical Review} only thanks to the authority of Rossi who personally assumed the responsibility for the assertions thereby contained. Thirty years had elapsed since the young Bruno Rossi had to request the authority of Werner Heisenberg\index{Heisenberg, Werner K.} to have the plausibility of his early scientific results validated.
AS\&E\index{American Science \& Engineering} took the lead for many years in developing X-ray astronomy\index{X-ray astronomy}, a  field which had a giant development also thanks to Riccardo Giacconi\index{Giacconi, Riccardo}  who further greatly contributed to the development of the techniques and constructed X-ray telescopes.

Nature makes science international in character and provides a strong basis for international cooperation in science. The space era, like the foundation of great laboratories such as CERN\index{CERN}  enlarged more and more this basis and opened new research lines in view of  the perspectives opened by the new-born space science. The vigorous expansion of space research at the beginning of the 1960s went in parallel with a general reorganization in the field of education and with pioneering researches in a variety of newly emerging fields of astrophysical character, whose interdisciplinary nature required the establishment of a dedicated center, which appeared crucial to the appropriate development of MIT's commitment to advanced research.  In April 1963 MIT established the Center for Space Research\index{Center for Space Research, MIT} (CSR), and in a Memorandum of Understanding signed in May with the National Aeronautics and Space Agency agreed to augment its teaching and research in space-related fields through the Center, while ``NASA\index{NASA} agreed to support a general research program within the Center, to continue certain large specific space research projects on campus, and to contribute heavily towards the cost of constructing a major new building on the campus to house this growing program.'' In 1965 the CSR transformed in the MIT Kavli Institute for Astrophysics and Space Research\index{Kavli Institute for Astrophysics and Space Research} (MKI), an interdepartmental center supporting research in space science and engineering, astronomy and astrophysics. 

Last but not least, the strong connection between Rossi and the beginning of space science in Italy emerges from his correspondence with Edoardo Amaldi\index{Amaldi, Edoardo}  who at the time was promoting an European Space Agency after his deep involvement in the foundation of CERN. Rossi was one of the roots of Italy in space, also because he inspired Giuseppe Occhialini\index{Occhialini, Giuseppe Stanislao}  who spent some time at MIT with his wife Connie Dilworth, who participated to the plasma probe experiment. Later Occhialini promoted space science in Milan, and became a well known leader in this field. Like Fermi, Rossi has always been a myth for young Italian physicists, always strongly supporting young people, and inviting them to work at MIT.

Two X-ray astronomy satellites, the {\it Bruno Rossi X-ray Timing Explorer}\index{Bruno Rossi X-ray Timing Explorer} (RXTE) and \textit{BeppoSAX}\index{BeppoSAX} are named after these two leaders of 20th century physics. This kind of satellites have revolutionized our understanding of X-ray Universe, particularly in helping to throw some light on the mysterious ``$\gamma$-ray bursts''\index{gamma-ray bursts}  the sudden intense flashes of $\gamma$-rays, which in a few seconds release an extremely large amount of energy, the equivalent of energy released by 1000 stars like the Sun over their entire lifetime. 

\section{Concluding remarks}

Bruno Rossi passed away in 1993. His active involvement in the research activity was completed in the first 1960s, though his experience and teaching continued to enrich the scientific community  in the course of the following  years. Indeed, the unrivalled importance of his work is strictly attached to what he offered to those he met, especially his disciples and co-workers-to be. Many present physicists have met him and experienced first-hand the vicissitudes of cosmic rays physics in the 1940s and 1950s, as well as the foundation of  space physics and X-ray astronomy since the end of 1950s. Thus, they consider Bruno Rossi a crucial figure in their personal history, and the {\it Rossi coincidence circuit} a  landmark in the history of 20th  century physics. 
 
The well known achievements of precise experimentation, of subtle instrument design, and of his outstanding manipulative skill are part and parcel of the craft of the experimental physicist, which at its highest became indeed with Rossi the Art of experimental physics. The high quality, both scientific and human, of the man and his style contributed to shape the myth of Bruno Rossi. It is the case that his fame would precede him, especially among the young who met him for the first time. His disciple and present Bruno Rossi Professor of Physics at MIT, Claude Canizares\index{Canizares, Claude R.}, beautifully described such atmosphere in a brief note (14 December, 1993), while still in grief for his mentor's death:\footnote{Copy of the document was kindly provided by Arlyn Hertz, Administrative Assistant to Jacqueline Hewitt,  Director of MIT Kavli Institute for Astrophysics and Space Research.}
 
 \begin{quotation}
 \small
 Long before he met me, Bruno spoke to me. Quietly. Patiently. In the solitude of my study, often late into the night, he revealed the wonders of optics and modern physics -- through his books [\dots] It was no surprise when, as a freshly minted postdoc [\dots] I heard one luminary after another heap praise upon this man, for his vast achievements, his immense influence as a visionary statesman of science, a teacher, a mentor [\dots]
 
Bruno's kindness was a virus, infecting anyone who came in contact with him. Despite all the profound lessons of physics that Bruno taught me before and after he met me, I learned still more from his style. Scientifically he combined two contradictory traits: on one hand, the daring of a fearless explorer, undaunted by the prospect of repeated forays into uncharted territory, daring to leap through the looking glass over and over again. On the other hand lay the caution and precision of a fine sculptor or master craftsman. Never straying beyond the certainty of his data, but playing it, like a virtuoso violinist for all it was worth [\dots]

Bruno's greeting, the way he said hello, was a metaphor for his personality [\dots] Bruno would swing his arm and hand sideways, in a short embracing arc evoking a figurative Latin hug, ending with a firm grasp of your hand, a conspiratorial and hearty shake, a smile that seemed to surge over his face, especially his eyes, and established -- without reservation -- the warmth and sincerity of his greeting, as if he had been waiting for you all day.

And in a sense, he had [\dots]
 \end{quotation}
 \normalsize

Rossi's intuition on the importance of exploiting  new technological windows to look at the universe with new eyes is  a fundamental key to understand the wide underlying theme which shaped the natural evolution of his scientific identity  from  a ``cosmic-ray physicist'' to a ``cosmic-ray astronomer''. His scientific path is strongly interlaced with cosmic rays as a key to understand first natural phenomena at a microscopic level, and later at a cosmic scale. The new field of astroparticle physics, which emerged at the intersection of particle physics, astronomy and cosmology, provides the proper perspective to look at  Rossi's scientific path. 

There certainly exists a profound unity in Rossi's work, guiding his  research line up to the culminating moment of his scientific career at the beginning of 1960s, when, as a  visionary  leader, he promoted two experiments  which inaugurated two new research fields. In 1961 his group at MIT made first {\it in situ} measurements of the interplanetary plasma around the Earth proving the existence of the solar wind and opening the study of the magnetospheres of other planets in the solar system.  In parallel Rossi  promoted the search for extra-solar sources of X rays, an idea which eventually led to the breakthrough experiment which in 1962 discovered Scorpius X-1. Cosmic X-ray astronomy today represents one the main instruments to investigate astrophysical processes and the nature of the celestial bodies generating them.\footnote{The remarkable development the field has gone through were especially due to the efforts of Riccardo Giacconi\index{Giacconi, Riccardo} ---at the time a young member of the first research {\it team}---  who later contributed to develop other detecting techniques and built telescopes to observe X rays coming from the space. Giacconi\index{Giacconi, Riccardo} was awarded  the Nobel Prize for Physics in 2003 ``for pioneering contributions to astrophysics, which have led to the discovery of cosmic X-ray sources.'' His mentor would have deserved that Nobel as well if it were not for his death ten years earlier. }

These achievements, which became the  final acts of his scientific life, represented the conclusion of a coherent project, undertaken in Arcetri thirty years before. During the same days, in August 1962, Rossi was writing the last pages of his well known volume {\it Cosmic Rays} where he recalled  Victor Hess'  discovery of the ``radiation from above'', and remarked how its 50th anniversary had come at a ``crucial moment for the physicists of cosmic rays and for the whole cosmic ray physics'' \cite{Rossi:1966aa}:

\begin{quotation}
\small
The interest in cosmic rays is certainly not waning; on the contrary, it is steadily growing. But cosmic-ray research has become such an integral part of many different scientific endeavors that it has almost ceased to exist as a separate and distinct branch of science. The ``cosmic-ray physicist,'' as a specialist, is becoming a figure of the past, while the nuclear physicist, the geophysicist, the astrophysicist, and the cosmologist are turning more and more to the study of cosmic rays for information of vital importance to the solution of their problems. It is quite possible that future historians of science will close the chapter on cosmic rays with the fiftieth anniversary of Hess's\index{Hess, Victor F.} discovery. However, they will undoubtedly note that in renouncing its individuality and merging with the main stream of science, cosmic-ray research continued to perform a vital role in advancing man's understanding of the physical world.
\end{quotation}
\normalsize 

We are now celebrating the centennial of the discovery of cosmic rays.  Some of these particles have such breathtaking energies, a hundred million times above that provided by terrestrial accelerators, that the questions about how can cosmic accelerators boost particles to these energies, and about what is the nature of the particles themselves, are still  open. The mystery of cosmic rays is nowadays tackled ---and maybe going to be solved--- by an interplay of sophisticated detectors for high-energy $\gamma$-rays, charged cosmic rays and neutrinos. 

 The long evolution leading to the present variety of research traditions highlights the great potentialities of a field which was shaped by  the scientific intelligence of these fathers of modern science and which is presently showing all its vitality in an era when high-energy physics with accelerators is certainly in a crucial and critical phase, even if full  of  great expectations, particularly deriving from the entering into action of the CERN Large Hadron Collider. 

The investigation of cosmic rays and of all kind of signals coming from the outer space, actually represents the necessary alternative to the high-energy physics with accelerators for the study of the Universe and for the understanding of the elementary processes which have originated the Universe itself. The wider and wider perspective deriving from these researches is largely confirming  the far-sighted vision of scientists like Rossi, who is certainly to be considered as one of the last ``natural philosophers'' of our epoch.

%\frontmatter%%%%%%%%%%%%%%%%%%%%%%%%%%%%%%%%%%%%%%%%%%%%%%%%%%%%%%

%\include{dedic}
%\include{foreword}
%\include{preface}
%\include{acknow}

%\tableofcontents

%\include{acronym}

%\mainmatter%%%%%%%%%%%%%%%%%%%%%%%%%%%%%%%%%%%%%%%%%%%%%%%%%%%%%%%
%\include{part}
%\include{GeneralIntroduction}

%\backmatter%%%%%%%%%%%%%%%%%%%%%%%%%%%%%%%%%%%%%%%%%%%%%%%%%%%%%%%
%\include{glossary}
%\include{solutions}
\bibliography{Bibliography}
%\printindex

%%%%%%%%%%%%%%%%%%%%%%%%%%%%%%%%%%%%%%%%%%%%%%%%%%%%%%%%%%%%%%%%%%%%%%

\end{document}